\documentclass[journal]{IEEEtran}
\IEEEoverridecommandlockouts
\usepackage{xcolor}        
\usepackage{graphicx}
\usepackage{amsmath, amssymb, amsfonts, amsthm}
\usepackage[caption=false, font=footnotesize]{subfig}
\usepackage{textcomp}
\usepackage{cite}
\usepackage{multirow}
\usepackage[ruled,vlined,linesnumbered]{algorithm2e}
\DeclareMathOperator*{\maximize}{maximize}
\DeclareMathOperator*{\minimize}{minimize}
\DeclareMathOperator{\diag}{diag}
\DeclareMathOperator{\tr}{tr}
\DeclareMathOperator{\tril}{tril}
\DeclareMathOperator{\re}{Re}

\newtheorem{remark}{Remark}

\newenvironment{rev}{}{}

\def\BibTeX{{\rm B\kern-.05em{\sc i\kern-.025em b}\kern-.08em
    T\kern-.1667em\lower.7ex\hbox{E}\kern-.125emX}}
\begin{document}

\title{Distributed Precoding for Cell-free Massive MIMO in O-RAN: A Multi-agent Deep Reinforcement Learning Framework\\ 
\thanks{This paper has been published in part in the \textit{Proceedings of IEEE International Conference on Communications (ICC)}, Montreal, Canada, June 2025 \cite{MIMO:MHSH1}.

The authors are with the Department of Electrical and Computer Engineering, The University of British Columbia, Vancouver, BC, V6T 1Z4, Canada (email: mhshokouhi@ece.ubc.ca, vincentw@ece.ubc.ca).}}

\author{\IEEEauthorblockN{Mohammad Hossein Shokouhi, \textit{Graduate Student Member, IEEE,} and Vincent W.S. Wong, \textit{Fellow, IEEE}}
}

\maketitle
\begin{abstract}
Cell-free massive multiple-input multiple-output (MIMO) is a key technology for next-generation wireless systems, where each user is served by multiple open radio units (O-RUs) collaboratively. The integration of cell-free massive MIMO within the open radio access network (O-RAN) architecture addresses the growing need for decentralized, scalable, and high-capacity networks that can support different use cases. Precoding is a crucial step in the operation of cell-free massive MIMO, where the O-RUs steer their beams towards the intended users while mitigating inter-user interference. Most of the existing precoding schemes for cell-free massive MIMO are either fully centralized or fully distributed. Centralized schemes are not scalable, whereas distributed schemes may lead to high inter-O-RU interference. In this paper, we propose a distributed and scalable precoding framework for cell-free massive MIMO that uses limited information exchange among precoding agents to mitigate interference. We formulate an optimization problem for precoding that maximizes the aggregate throughput while guaranteeing the minimum data rate requirements of users. The formulated problem is nonconvex. We leverage the O-RAN architecture and propose a multi-timescale framework that combines multi-agent deep reinforcement learning (DRL) with expert insights from an iterative algorithm to determine the precoding matrices efficiently. We conduct simulations and compare the proposed framework with the centralized precoding and distributed precoding methods for different numbers of O-RUs, users, and transmit antennas. The results show that the proposed framework achieves a higher aggregate throughput than the distributed regularized zero-forcing (D-RZF) scheme and the weighted minimum mean square error (WMMSE) algorithm. When compared with the centralized regularized zero-forcing (C-RZF) scheme, the proposed framework achieves similar aggregate throughput performance but with a lower signaling overhead. We also demonstrate that the proposed framework can dynamically adapt to changes in the minimum data rate requirements.
\end{abstract}
\section{Introduction}
\begin{figure}[t]
\center{\includegraphics[height=98mm]{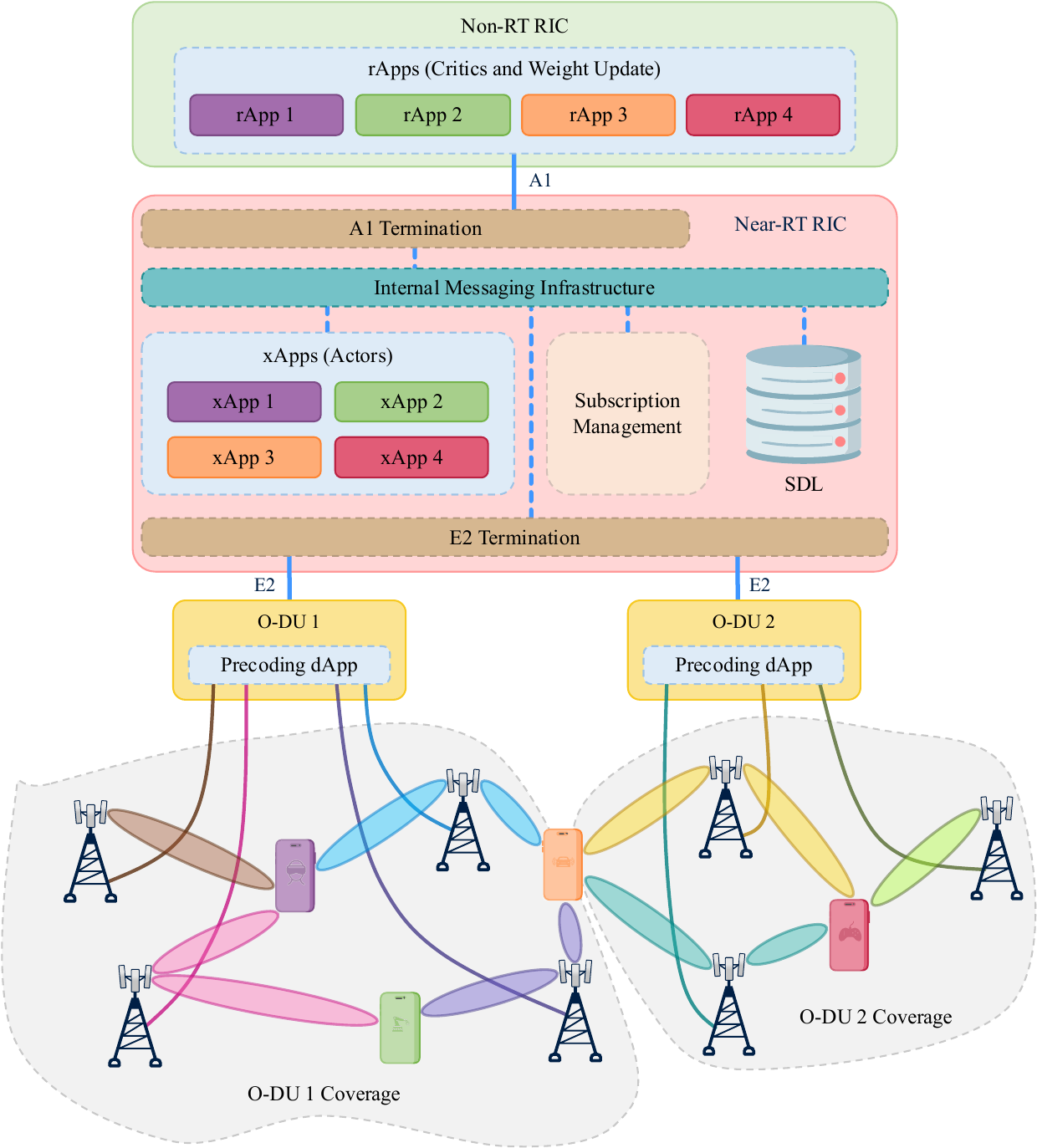}}
\caption{The considered cell-free massive MIMO system in O-RAN. The rApps handle DNN training and provide feedback to the DRL agents. The xApps obtain the intermediate variables for the users. The dApps determine the precoding matrices in a distributed manner at each RT loop.}
\label{fig:system_model}
\end{figure}

With the emergence of technologies such as cell-free massive multiple-input multiple-output (MIMO), effective wireless resource management demands solutions that provide access to data and analytics and enable data-driven optimization. To address this need, the open radio access network (O-RAN) paradigm has been proposed in the literature \cite{survey:O-RAN1}. O-RAN promotes a virtualized radio access network (RAN) in which disaggregated components are interconnected via open interfaces and are optimized by intelligent controllers. The O-RAN architecture splits the functions of the next-generation node B (gNB) into three components \cite{O_RAN_description}: the open central unit (O-CU), the open distributed unit (O-DU), and the open radio unit (O-RU). Specifically, O-RAN adopts the functional split option 7-2x defined by the Third Generation Partnership Project (3GPP) \cite{O_RAN_split, 3GPP_split_options}. Under this split option, functions such as cyclic prefix and inverse fast Fourier transform are handled by the O-RUs. Other functions such as precoding and modulation, along with medium access control and radio link control operations, are performed by the O-DUs. Higher-layer functions are executed at the O-CUs. These components are deployed hierarchically, where each O-DU serves multiple O-RUs \cite{O_RAN_description}, as shown in Fig. \ref{fig:system_model}. 

O-RAN also features two RAN intelligent controllers (RICs) that provide a centralized view of the network and enable the control and optimization of RAN at different timescales: the near-real-time (near-RT) RIC, which manages the network in a near-RT (10 ms to 1 sec) timescale, and the non-RT RIC, which operates at the non-RT (over 1 sec) timescale. The near-RT RIC collects data from the O-DUs and O-CUs via the E2 interface and leverages machine learning (ML) algorithms to select the actions. It hosts microservices called xApps that support optimization routines and ML workflows. The near-RT RIC also includes the shared data layer (SDL), which is a centralized database that enables xApps to store, retrieve, and share data through standardized application programming interfaces (APIs) \cite{Near_RT_RIC_Architecture}. The non-RT RIC hosts rApps that train ML models and generate policies, which are sent to the near-RT RIC via the A1 interface \cite{O_RAN_description}. Furthermore, dApps are introduced in \cite{melodia:dApp_full} that run on O-DUs to support RT (below 1 ms) control loops in O-RAN. 

Recent works have proposed various xApps and rApps to enable data-driven closed-loop control within O-RAN.
In \cite{melodia:xApp-scheduler}, an xApp aims to dynamically update the priority coefficients of users in a proportional fair scheduler in order to guarantee the minimum data rate for each user. 
In \cite{eMBB-URLLC:dist1}, a deep reinforcement learning (DRL) agent, which is deployed as an xApp in the near-RT RIC, is assigned to each O-RU. The goal is to support power control and radio resource allocation.

In recent years, cell-free massive MIMO has emerged as a promising wireless technology, where each user can be served by multiple O-RUs \cite{CF-MMIMO}. Compared with the conventional wireless cellular architectures, cell-free massive MIMO architecture can achieve more uniform data rates across the coverage area due to the macro-diversity gain offered by distributed O-RUs. 
The integration of cell-free massive MIMO in O-RAN enables decentralized, scalable, and intelligent network management.
In \cite{bjornson:O-RAN1}, an optimization problem is formulated to minimize the power consumption of the RAN nodes in a cell-free O-RAN by jointly optimizing the radio, optical fronthaul, and cloud processing resources.
In \cite{MIMO:pilot1}, a multi-agent DRL algorithm is proposed in the near-RT RIC for pilot sequence assignment to the users.

Precoding is a crucial step in the operation of massive MIMO systems, where the O-RUs steer their signal beams towards the intended users in order to enhance the signal strength, reduce interference, and improve the energy efficiency.
Precoding has been extensively studied in the cellular massive MIMO literature.
In \cite{MIMO1}, the problem of maximizing the aggregate throughput subject to the O-RU transmit power constraint is formulated as a weighted minimum mean square error (WMMSE) problem with the weight, receive filter, and precoding matrices as the optimization variables. The problem is solved iteratively using block coordinate descent (BCD). The precoding subproblem is decoupled across O-RUs and is solved in closed form.
\begin{rev}
However, the original WMMSE algorithm \cite{MIMO1} is not applicable to the precoding problem in cell-free massive MIMO systems due to their distinct architecture, where each user is served by multiple O-RUs.
\end{rev}
In \cite{WMMSE:reduced}, a variant of the WMMSE algorithm is proposed for cell-free multiple-input single-output (MISO) networks. 
In \cite{WMMSE:non_coherent}, a WMMSE-based beamforming algorithm is proposed for cell-free networks with noncoherent joint transmission, where each O-RU transmits an independent data stream to a user without requiring phase synchronization. 
\begin{rev}
The aforementioned works \cite{MIMO1, WMMSE:reduced, WMMSE:non_coherent} rely on iterative optimization algorithms to determine the precoding matrices, which may encounter scalability issues in large-scale deployments with many users and O-RUs.
\end{rev}

\begin{rev}
To address the scalability issue, some recent works use ML to reduce the computational complexity of the WMMSE algorithm.
\end{rev}
In \cite{MIMO:WMMSE-unfolding}, the WMMSE algorithm is modeled using a deep neural network (DNN) that uses trainable parameters to approximate those high-complexity operations such as matrix inversion. After the DNN has been trained, it can achieve performance close to WMMSE.
In \cite{MIMO:SAC_QoS}, the soft actor-critic (SAC) DRL algorithm is used to determine the priority weights of users based on the queue length. These weights are then provided to the WMMSE algorithm for precoding.
In \cite{MIMO:E2E1}, a DNN, which is deployed at the O-RU in time division duplexing mode, uses the received uplink pilot signals to estimate the effective channel and determine the weight matrix and power allocation coefficients. These outputs are then used in the update equation of the WMMSE algorithm to determine the precoding matrices.
A similar approach is proposed in \cite{MIMO:E2E2} for the frequency division duplexing mode.
The multi-cell massive MIMO scenario is considered in \cite{MIMO:precoding_DRL2}, where a DRL agent at each O-RU uses local channel state information (CSI) along with historical information from other O-RUs to determine the power allocation and weight coefficients. 
\begin{rev}
The aforementioned works \cite{MIMO:WMMSE-unfolding, MIMO:E2E2, MIMO:E2E1, MIMO:precoding_DRL2, MIMO:SAC_QoS} consider cellular massive MIMO systems and use the WMMSE algorithm \cite{MIMO1} to determine the precoding matrices. As discussed earlier, the original WMMSE algorithm \cite{MIMO1} is not applicable to the precoding problem in cell-free massive MIMO systems due to the inherent coupling of precoding matrices across O-RUs.
\end{rev}

\begin{rev}
Recently, some DRL-based approaches have been proposed to directly determine the precoding matrices from CSI in an end-to-end manner, without relying on optimization algorithms.
\end{rev}
In \cite{MIMO:precoding_DRL3}, two O-RUs in a cellular massive MIMO network cooperatively determine the precoding matrices using a multi-agent DRL algorithm.
In \cite{MIMO:precoding_DRL4}, a centralized DRL algorithm uses the signal-to-interference-plus-noise ratios (SINRs) of the users to determine the precoding matrices in a cell-free network with the goal of maximizing the energy efficiency. 
In the conference version of this work \cite{MIMO:MHSH1}, we proposed a multi-agent actor-critic DRL algorithm for precoding. An actor, which is deployed as a dApp, is assigned to each O-RU. It uses the local CSI to determine its precoding matrix. A centralized critic at the near-RT RIC uses the states and actions information from all the actors to estimate the action-value function, which is used by the actors to update their policies.
\begin{rev}
The aforementioned works \cite{MIMO:precoding_DRL3, MIMO:precoding_DRL4, MIMO:MHSH1} use DRL algorithms to directly determine the precoding matrices. Recent works in \cite{MIMO:E2E1, MIMO:E2E2} show that this approach may lead to convergence issue in dense environments with many O-RUs and users due to the high-dimensional state and action spaces.
\end{rev}

\begin{rev}
In summary, the existing optimization-based precoding algorithms are primarily designed for cellular massive MIMO and not directly applicable to cell-free massive MIMO. Meanwhile, DRL-based approaches may suffer from convergence issues in dense deployments. Moreover, these schemes focus only on maximizing the aggregate throughput without considering the minimum data rate requirements of the users.
\end{rev}

\begin{rev}
In this paper, we propose a distributed precoding framework for cell-free massive MIMO. First, we propose a precoding optimization algorithm for cell-free massive MIMO with coherent joint transmission. We then use the communication-domain expert knowledge from the optimization algorithm and propose a multi-agent DRL framework that operates across different timescales to determine the precoding matrices efficiently. Here, the expert knowledge corresponds to using the insights from optimization-based methods to guide and improve the performance of data-driven models. 
\end{rev}
The main contributions of this paper are as follows:
\begin{itemize}
    \item We formulate an optimization problem for maximizing the aggregate throughput while guaranteeing the minimum data rate requirements of the users. 
    The formulated problem is nonconvex. We reformulate it as an equivalent WMMSE problem. Since the O-RUs collaboratively serve each user, their precoding matrices are coupled and cannot be determined independently. We apply the BCD approach to determine the precoding matrices iteratively.
    \item The iterative algorithm is computationally intensive as it involves multiple matrix inversions and requires many iterations to converge. To address this issue, we propose a multi-agent DRL framework, where a DRL agent is assigned to each user. Instead of directly learning the high-dimensional precoding matrices, each agent learns to determine a set of low-dimensional intermediate variables. These variables are used in the final update equation of the iterative algorithm to obtain the precoding matrices.
    \item We utilize the hierarchical architecture of O-RAN and decompose the proposed precoding framework into multiple stages. Each stage has a different timescale. Different RAN nodes are assigned to handle different stages. This reduces the computational load at the O-DUs and enables real-time precoding. At the non-RT RIC, rApps are responsible for training the DNNs and updating the parameters of the DRL agents. The near-RT RIC hosts the xApps, which are used to determine the user-specific intermediate variables. The outputs are then forwarded to the dApps at the O-DUs to determine the final precoding matrices in a distributed manner at each RT loop. The proposed framework uses limited information exchange among O-DUs to mitigate interference.
    \item Extensive simulations are carried out under different numbers of users, O-RUs, and transmit antennas. Results show that the proposed framework can achieve up to $24.4\%$ and $35.8\%$, and $50.2\%$ higher aggregate throughput when compared with distributed regularized zero-forcing (D-RZF), the WMMSE algorithm, and the end-to-end DRL algorithm, respectively. The proposed framework achieves similar performance when compared with the centralized regularized zero-forcing (C-RZF) scheme. We also evaluate the signaling overhead on the E2 interface for different numbers of users, cluster sizes, and transmit antennas. Due to its distributed nature, the proposed framework reduces the load on the E2 interface by up to $99.81\%$ when compared with the C-RZF scheme. Results from the computational complexity analysis show that the proposed framework scales efficiently with the number of users and O-RUs. Furthermore, the proposed framework can dynamically adapt to the changes in the data rate requirements. Finally, we evaluate the performance of the proposed framework under imperfect CSI and show that it achieves higher aggregate throughput than the centralized method in the presence of severe channel estimation error.
\end{itemize}

This paper is organized as follows. Section \ref{sec:model} presents the system model, the problem formulation, and an iterative algorithm. In Section \ref{sec:algorithm}, we propose a multi-agent DRL framework to determine the precoding matrices with low computational complexity. Performance evaluation is provided in Section \ref{sec:eval}. The conclusion is given in Section \ref{sec:conclusion}.

{\it Notations}: In this paper, $\mathbb{C}$ and $\mathbb{R}$ denote the set of complex and real numbers, respectively. Boldface uppercase letters (e.g., $\mathbf{X}$) represent matrices, while boldface lowercase letters (e.g., $\mathbf{x}$) represent vectors. The $N \times N$ identity matrix is denoted by $\mathbf{I}_N$. $(\cdot)^\top$ and $(\cdot)^\mathrm{H}$ denote the transpose and conjugate transpose of a vector or matrix. For a matrix, $\tr(\cdot)$ and $\det(\cdot)$ denote the trace and determinant, respectively, and $\tril(\cdot)$ denotes its lower-triangular part with entries above the main diagonal set to zero. $\diag(\cdot)$ returns the vector of diagonal elements of a square matrix. The notation $[\cdot]_{i,j}$ refers to the element in row $i$ and column $j$ of a matrix, and $\re(\cdot)$ denotes the real part of a complex matrix.

\begin{figure}[t]
\center{\includegraphics[width=\linewidth]{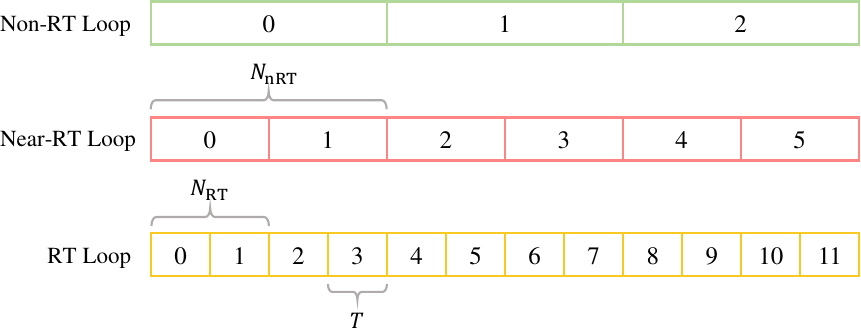}}
\caption{The timescales of control loops within O-RAN. A non-RT loop occurs once every $N_\textrm{nRT}$ near-RT loops. A near-RT loop occurs once every $N_\textrm{RT}$ RT loops. Each RT loop has a duration of $T$ seconds.}
\label{fig:timescales}
\end{figure}

\section{System Model and Problem Formulation} \label{sec:model}
We consider the downlink operation of a cell-free massive MIMO system within the O-RAN architecture, illustrated in Fig. \ref{fig:system_model}. There are $K$ users. They are served on the same time-frequency resources via spatial multiplexing. Let $\mathcal{K}=\{1,2,\ldots,K\}$ denote the set of users. The O-RAN consists of $L$ O-RUs denoted by set $\mathcal{L}=\{1,2,\ldots,L\}$ and $U$ O-DUs denoted by set $\mathcal{U}=\{1,2,\ldots,U\}$. Each O-RU has $N_\textrm{t}$ transmit antennas. Each user equipment has $N_\textrm{r}$ receive antennas. Each O-DU $u \in \mathcal{U}$ serves a subset of O-RUs $\mathcal{L}^\textrm{DU}_u$ using open fronthaul (O-FH) links. As shown in Fig. \ref{fig:timescales}, a near-RT loop occurs once every $N_\textrm{RT}$ RT loops. A non-RT loop occurs once every $N_\textrm{nRT}$ near-RT loops. Each RT loop has a duration of $T$ seconds. 
Let $t$ denote the current RT loop index and $n$ denote the current near-RT loop index. Near-RT loop $n$ begins at RT loop $t=nN_\textrm{RT}$.

The downlink channel matrix $\mathbf{H}_{k,l}(t) \in \mathbb{C}^{N_\textrm{r} \times N_\textrm{t}}$ between O-RU $l \in \mathcal{L}$ and user $k \in \mathcal{K}$ during RT loop $t$ is given by
\begin{equation}\label{eq:channel_def}
    \mathbf{H}_{k,l}(t) = \sqrt{\beta_{k,l}} \mathbf{G}_{k,l}(t),
\end{equation}
where $\beta_{k,l}$ denotes the large-scale fading coefficient between user $k$ and O-RU $l$. $\mathbf{G}_{k,l}(t)\in \mathbb{C}^{N_\textrm{r} \times N_\textrm{t}}$ is the small-scale fading matrix, which evolves according to a first-order Gauss–Markov process, given by
\begin{equation}\label{eq:small_fading}
    \mathbf{G}_{k,l}(t) = \epsilon_k \mathbf{G}_{k,l}(t-1) + \sqrt{(1-\epsilon_k^2)} \mathbf{\Omega}_{k,l}(t),
\end{equation}
where the entries of the matrix $\mathbf{\Omega}_{k,l}(t)\in \mathbb{C}^{N_\textrm{r} \times N_\textrm{t}}$ are independent and identically distributed (i.i.d.) random variables following the complex Gaussian distribution with zero mean and unit variance, i.e., $\mathcal{CN}(0,1)$. $\epsilon_k = J_0 \left(2 \pi \frac{v_k}{c} f_\textrm{c} T\right)$ is the temporal correlation coefficient for user $k$ \cite{MIMO:pilot1} and $J_0(.)$ is the Bessel function of the first kind of order zero. $v_k$, $c$, and $f_\textrm{c}$ denote the velocity of user $k$, the speed of light, and the carrier frequency, respectively. For brevity, we omit the loop index from the equations throughout the rest of this section.

In cell-free massive MIMO, each user is served by a subset of O-RUs selected on a user-centric basis according to the user’s channel conditions. 
Let $\mathcal{K}_l \subset \mathcal{K}$ and $K_l$ denote the subset and the number of users served by O-RU $l$, respectively. Furthermore, let $\mathcal{L}^\textrm{UE}_k \subset \mathcal{L}$ and $L^\textrm{UE}_k=L^\textrm{UE}$ denote the subset and the number of O-RUs that serve user $k$, respectively. 

Let $\mathbf{V}_{k,l} \in \mathbb{C}^{N_\textrm{t} \times N_\textrm{s}}$ denote the precoding matrix at O-RU $l$ for data transmission to user $k \in \mathcal{K}_l$, where $N_\textrm{s}=\min(N_\textrm{t}, N_\textrm{r})$ is the number of data streams. 
The downlink signal transmitted by O-RU $l \in \mathcal{L}$ can be expressed as
\begin{equation}\label{eq:x_l}
    \mathbf{x}_l = \sum_{k \in \mathcal{K}_l} \mathbf{V}_{k,l} \mathbf{s}_k,
\end{equation}
where $\mathbf{s}_k \in \mathbb{C}^{N_\textrm{s}}$ is the data symbol vector for user $k$ and $\mathbb{E}\left[\mathbf{s}_k \mathbf{s}_k^\textrm{H}\right] = \mathbf{I}_{N_\textrm{s}}$. The signal received by user $k$ is
\begin{equation}\label{eq:y_k}
    \mathbf{y}_k = \sum_{l \in \mathcal{L}} \mathbf{H}_{k,l} \mathbf{x}_l + \mathbf{n}_k,
\end{equation}
where $\mathbf{n}_k \in \mathbb{C}^{N_\textrm{r}}$ is the additive white Gaussian noise vector that follows a complex Gaussian distribution with zero mean and covariance matrix $\sigma^2 \mathbf{I}_{N_\textrm{r}}$, i.e., $\mathcal{CN}\left(0,\sigma^2 \mathbf{I}_{N_\textrm{r}}\right)$. By substituting \eqref{eq:x_l} into \eqref{eq:y_k}, we obtain
\begin{equation}
    \mathbf{y}_k = \underbrace{\sum_{l \in \mathcal{L}^\textrm{UE}_k} \mathbf{H}_{k,l} \mathbf{V}_{k,l} \mathbf{s}_k}_{\text{Desired signal}} + \underbrace{\sum_{l \in \mathcal{L}} \sum_{i \in \mathcal{K}_l \setminus \{k\}} \mathbf{H}_{k,l} \mathbf{V}_{i,l} \mathbf{s}_i}_{\text{Inter-user interference}} \;+\; \mathbf{n}_k.
\end{equation}
We assume that the signals for different users are independent of each other. User $k$ applies the receive filter $\mathbf{U}_k \in \mathbb{C}^{N_\textrm{r}\times N_\textrm{s}}$ to extract its intended signal from $\mathbf{y}_k$ as
\begin{equation}
    \hat{\mathbf{s}}_k = \mathbf{U}_k^\textrm{H} \mathbf{y}_k.
\end{equation}
The achievable data rate of user $k \in \mathcal{K}$ can be written as
\begin{equation} \label{eq:eMBB_rate}
   r_{k} = \log_2 \det \left(\mathbf{I}_{N_\textrm{r}} + \mathbf{\Gamma}_{k}\right),
\end{equation}
where $\mathbf{\Gamma}_{k} \in \mathbb{C}^{N_\textrm{r} \times N_\textrm{r}}$ is the SINR matrix of user $k$, given by \cite{data_rate2:wong-mehrabian}
\begin{equation}\label{eq:SINR}
    \mathbf{\Gamma}_{k} = \mathbf{\Xi}_{k,k}\mathbf{\Xi}_{k,k}^\textrm{H} \left(\sum_{i \in \mathcal{K} \setminus \{k\}} \mathbf{\Xi}_{k,i}\mathbf{\Xi}_{k,i}^\textrm{H} + \sigma^2 \mathbf{I}_{N_\textrm{r}} \right)^{-1}.
\end{equation}
In \eqref{eq:SINR}, $\mathbf{\Xi}_{k,i} \in \mathbb{C}^{N_\textrm{r} \times N_\textrm{s}}$ denotes the effective channel matrix for user $k$ if $i=k$, and the effective interference matrix from user $i$ to user $k$ otherwise. It can be determined by
\begin{equation}\label{eq:effective_matrix}
    \mathbf{\Xi}_{k,i} = \sum_{l \in \mathcal{L}^\textrm{UE}_i} \mathbf{H}_{k,l} \mathbf{V}_{i,l}.
\end{equation}

We aim to maximize the aggregate throughput while guaranteeing the minimum data rate requirements of the users. The precoding optimization problem can be formulated as
\begin{subequations} \label{eq:optimization}
\begin{alignat}{2}
&\maximize\limits_{\substack{\mathbf{V}_{k,l},\\ k \in \mathcal{K}_l, l \in \mathcal{L}}} & &\quad\sum_{k \in \mathcal{K}} r_{k} \label{eq:obj}\\
&\textrm{subject to} & &\quad r_{k} \geq R_{k}^{\textrm{min}}, \; k \in \mathcal{K} \label{eq:cons1}\\
& & &\quad \sum_{k \in \mathcal{K}_l} \tr\left(\mathbf{V}_{k,l}\mathbf{V}_{k,l}^\textrm{H}\right) \leq P^{\textrm{max}}, \; l \in \mathcal{L}, \label{eq:cons2}
\end{alignat}
\end{subequations}
where $R_{k}^{\textrm{min}}$ denotes the minimum data rate requirement of user $k \in \mathcal{K}$ and $P^{\textrm{max}}$ denotes the maximum transmit power at each O-RU.
To achieve real-time precoding, problem \eqref{eq:optimization} must be solved in each RT loop. 
However, the objective function \eqref{eq:obj} and constraint \eqref{eq:cons1} are nonconvex.
In the following subsection, we propose a distributed algorithm to solve problem \eqref{eq:optimization} in an iterative manner.

\subsection{Iterative Algorithm}\label{subsec:optimal_structure}
We introduce the set of Lagrange multipliers $\left\{\mu_k:k \in \mathcal{K}\right\}$ to incorporate constraint \eqref{eq:cons1} into the objective function. Constraint \eqref{eq:cons2} remains as an explicit constraint in the dual problem. The partial Lagrange dual function is
\begin{equation}\label{eq:dual_function}
    g(\boldsymbol{\mu}) = \sup_{\mathbf{V} \in \mathcal{D}} \sum_{k \in \mathcal{K}} \left(r_{k} + \mu_k \left(r_{k} - R_{k}^{\textrm{min}}\right)\right),
\end{equation}
where $\mathbf{V} = \left[\mathbf{V}_{k,l},\;\forall k \in \mathcal{K}, l \in \mathcal{L}\right] \in \mathbb{C}^{LN_\textrm{t} \times KN_\textrm{s}}$ is the stacked precoding matrix, $\boldsymbol{\mu}= \left[\mu_k,\;\forall k \in \mathcal{K}\right]^\top$ is the vector of Lagrange multipliers, and $\mathcal{D}=\left\{\mathbf{V}: \sum_{k \in \mathcal{K}_l} \tr\left(\mathbf{V}_{k,l}\mathbf{V}_{k,l}^\textrm{H}\right) \leq P^{\textrm{max}}, l \in \mathcal{L}\right\}$. 
To obtain \eqref{eq:dual_function}, we need to solve the inner supremum over $\mathbf{V}$. Since the $\mu_k R_{k}^{\textrm{min}}$ in \eqref{eq:dual_function} does not depend on $\mathbf{V}$, they can be omitted when solving this subproblem. Thus, we have
\begin{align}
&\maximize\limits_{\substack{\mathbf{V}_{k,l},\\ k \in \mathcal{K}_l, l \in \mathcal{L}}} \quad\sum_{k \in \mathcal{K}} \omega_k r_{k} \label{eq:inner_supremum}\\
&\textrm{subject to constraint \eqref{eq:cons2}}, \notag
\end{align}
where $\omega_k=1+\mu_k$. 
The Lagrange dual problem can be expressed as
\begin{subequations} \label{eq:dual_problem}
\begin{alignat}{2}
&\inf_{\mu_k, k \in \mathcal{K}} & &\quad g(\boldsymbol{\mu}) \label{eq:dual_obj}\\
&\textrm{subject to} & &\quad \mu_k \geq 0, \; k \in \mathcal{K}. \label{eq:dual_cons}
\end{alignat}
\end{subequations}
We use the dual gradient ascent method to iteratively solve subproblems \eqref{eq:inner_supremum} and \eqref{eq:dual_problem} for $\mathbf{V}$ and $\boldsymbol{\mu}$, respectively. It has been proven in \cite{MIMO1} that the WMMSE problem formulated as 
\begin{align}
&\minimize_{\mathbf{W}, \mathbf{U}, \mathbf{V}} \quad\sum_{k \in \mathcal{K}} \omega_{k}\left(\tr(\mathbf{W}_k \mathbf{E}_k) - \log_2 \det (\mathbf{W}_k)\right) \label{eq:WMMSE_problem}\\
&\textrm{subject to constraint \eqref{eq:cons2}}, \notag
\end{align} 
is equivalent to the weighted sum-rate maximization problem \eqref{eq:inner_supremum}, and both problems yield the same optimal solution $\mathbf{V}^*$. In \eqref{eq:WMMSE_problem}, $\mathbf{W}_k \in \mathbb{C}^{N_\textrm{s} \times N_\textrm{s}}$ is an auxiliary variable that denotes the weight matrix of user $k$. $\mathbf{U}=\left[\mathbf{U}_1\ldots\mathbf{U}_K\right] \in \mathbb{C}^{N_\textrm{r} \times KN_\textrm{s}}$ and $\mathbf{W}=\left[\mathbf{W}_1\ldots\mathbf{W}_K\right] \in \mathbb{C}^{N_\textrm{s} \times KN_\textrm{s}}$ are the stacked receive filter and weight matrices, respectively. Furthermore, $\mathbf{E}_k \in \mathbb{C}^{N_\textrm{s} \times N_\textrm{s}}$ denotes the mean squared error (MSE) matrix for user $k$, which is given by
\begin{align}\label{eq:MSE}
    \mathbf{E}_k &\triangleq \mathbb{E}_{\mathbf{s},\mathbf{n}}\!\left[\!\left(\hat{\mathbf{s}}_{k} - \mathbf{s}_k\right)\!\left(\hat{\mathbf{s}}_{k} - \mathbf{s}_k\right)^\textrm{H}\right] \nonumber \\
    &= \left(\mathbf{I}_{N_\textrm{s}} - \mathbf{U}_k^\textrm{H} \mathbf{\Xi}_{k,k} \right) 
       \left(\mathbf{I}_{N_\textrm{s}} - \mathbf{U}_k^\textrm{H} \mathbf{\Xi}_{k,k} \right)^\textrm{H} \nonumber \\
    &\quad + \sum_{i \in \mathcal{K} \setminus \{k\}} 
       \mathbf{U}_k^\textrm{H} \mathbf{\Xi}_{k,i} \mathbf{\Xi}_{k,i}^\textrm{H} \mathbf{U}_k 
       + \sigma^2 \mathbf{U}_k^\textrm{H} \mathbf{U}_k.
\end{align}
Problem \eqref{eq:WMMSE_problem} is convex with respect to each of the individual optimization variables $\mathbf{W}, \mathbf{U}, \textrm{ and } \mathbf{V}$.
By fixing $\mathbf{U}$ and $\mathbf{V}$, the optimal $\mathbf{W}_k^*$ for user $k$ can be obtained using the first-order optimality condition as \cite{MIMO1}
\begin{equation}\label{eq:W_WMMSE}
    \mathbf{W}_k^* = \mathbf{E}_k^{-1}, \quad k \in \mathcal{K}.
\end{equation}
Moreover, by fixing $\mathbf{V}$, the optimal $\mathbf{U}_k^*$ can be determined as
\begin{equation}\label{eq:U_WMMSE}
    \mathbf{U}_k^* = \mathbf{J}_k^{-1} \mathbf{\Xi}_{k,k},
\end{equation}
where $\mathbf{J}_k = \sum_{i \in \mathcal{K}} \mathbf{\Xi}_{k,i} \mathbf{\Xi}_{k,i}^\textrm{H} + \sigma^2 \mathbf{I}_{N_\textrm{r}}$ is the covariance matrix of the signal received by user $k$. Finally, by holding $\mathbf{W}$ and $\mathbf{U}$ fixed and optimizing for $\mathbf{V}$, the following optimization problem can be formulated:
\begin{align}
&\minimize_{\substack{\mathbf{V}_{k,l},\\ k \in \mathcal{K}_l, l \in \mathcal{L}}} \quad\sum_{k \in \mathcal{K}} \omega_{k}\tr\left[\sum_{i \in \mathcal{K}}\mathbf{\Xi}_{k,i}^\textrm{H} \mathbf{X}_k \mathbf{\Xi}_{k,i}-2\re\left\{\mathbf{Y}_k \mathbf{\Xi}_{k,k}\right\}\right] \label{eq:WMMSE_V}\\
&\textrm{subject to constraint \eqref{eq:cons2}}, \notag
\end{align}
where $\mathbf{X}_k \in \mathbb{C}^{N_\textrm{r} \times N_\textrm{r}}$ and $\mathbf{Y}_k \in \mathbb{C}^{N_\textrm{s} \times N_\textrm{r}}$ are defined as
\begin{equation}\label{eq:A_definition}
    \mathbf{X}_k \triangleq \mathbf{U}_k \mathbf{W}_k \mathbf{U}_k^\textrm{H},
\end{equation}
\begin{equation}
    \mathbf{Y}_k \triangleq \mathbf{W}_k \mathbf{U}_k^\textrm{H}.
\end{equation}
The details of reformulating problem \eqref{eq:WMMSE_problem} into problem \eqref{eq:WMMSE_V} are presented in Appendix A.
\begin{remark} \label{remark:no_WMMSE}
    Unlike the original WMMSE algorithm proposed in \cite{MIMO1}, problem \eqref{eq:WMMSE_V} cannot be decoupled across O-RUs. This is because in cell-free massive MIMO with coherent joint transmission, multiple O-RUs collaboratively serve each user. Their precoding matrices are coupled and cannot be determined independently.
\end{remark}
Remark \ref{remark:no_WMMSE} motivates us to use the BCD method \cite{nonlinear_prog} to iteratively determine the precoding matrices for each O-RU by holding the other variables fixed. By using BCD, the optimization problem for each O-RU $l \in \mathcal{L}$ is formulated as
\begin{subequations} \label{eq:WMMSE_ORU}
\begin{align}
&\begin{aligned}\minimize_{\substack{\mathbf{V}_{k,l},\\ k \in \mathcal{K}_l}}\quad2\sum_{i \in \mathcal{K}}\omega_i\sum_{k \in \mathcal{K}_l}\re\left\{\tr\left[\mathbf{Z}_{i,k,l}^\textrm{H}\mathbf{X}_i\mathbf{H}_{i,l}\mathbf{V}_{k,l}\right]\right\}\\
+\;\sum_{i \in \mathcal{K}} \omega_i\sum_{k \in \mathcal{K}_l}\tr\left[\mathbf{X}_i \mathbf{H}_{i,l} \mathbf{V}_{k,l} \mathbf{V}_{k,l}^\textrm{H} \mathbf{H}_{i,l}^\textrm{H}\right]\\
-\;2\sum_{k \in \mathcal{K}_l} \omega_k \re\left\{\tr\left[\mathbf{Y}_k \mathbf{H}_{k,l} \mathbf{V}_{k,l}\right]\right\}\end{aligned} \label{eq:WMMSE_ORU_obj}\\
&\textrm{subject to} \quad \sum_{k \in \mathcal{K}_l} \tr\left(\mathbf{V}_{k,l}\mathbf{V}_{k,l}^\textrm{H}\right) \leq P^{\textrm{max}}, \label{eq:WMMSE_ORU_const}
\end{align}
\end{subequations}
where $\mathbf{Z}_{i,k,l} \in \mathbb{C}^{N_\textrm{r} \times N_\textrm{s}}$ is defined as
\begin{equation}\label{eq:C_term}
    \mathbf{Z}_{i,k,l} \triangleq \sum_{j \in \mathcal{L}^\textrm{UE}_k \setminus\{l\}} \mathbf{H}_{i,j} \mathbf{V}_{k,j}.
\end{equation}
The details of formulating subproblem \eqref{eq:WMMSE_ORU} based on problem \eqref{eq:WMMSE_V} are presented in Appendix B. The objective function in problem \eqref{eq:WMMSE_ORU} is a quadratic function, which is convex with respect to $\mathbf{V}_{k,l}$. Thus, using the Lagrange multipliers method \cite{convex}, the closed-form solution can be obtained as
\begin{align}\label{eq:V_WMMSE}
        \mathbf{V}_{k,l}^* = &\left(\sum_{i \in \mathcal{K}}\omega_i \mathbf{H}_{i,l}^\textrm{H} \mathbf{X}_i \mathbf{H}_{i,l} + \xi_l \mathbf{I}_{N_\textrm{t}}\right)^{-1} \nonumber\\ &\left(\omega_k\mathbf{H}_{k,l}^\textrm{H} \mathbf{Y}_k^\textrm{H} - \sum_{i \in \mathcal{K}} \omega_i \mathbf{H}_{i,l}^\textrm{H} \mathbf{X}_i^\textrm{H} \mathbf{Z}_{i,k,l}\right),
\end{align}
where $\xi_l \geq 0$ is a Lagrange multiplier. It can be determined using the complementary slackness condition of constraint \eqref{eq:WMMSE_ORU_const}, given by
\begin{equation}\label{eq:comp_slackness}
    \xi_l \left( \sum_{k \in \mathcal{K}_l} \tr \left( \mathbf{V}_{k,l} \mathbf{V}_{k,l}^\textrm{H} \right) - P^{\textrm{max}} \right) = 0.
\end{equation}
Let $\mathbf{V}_{k,l}(\xi_l)$ denote the right-hand side of \eqref{eq:V_WMMSE}. According to \eqref{eq:comp_slackness}, if $\sum_{k \in \mathcal{K}_l} \tr \left( \mathbf{V}_{k,l}(0) \mathbf{V}_{k,l}^\textrm{H}(0) \right) \leq P^{\textrm{max}}$, then $\mathbf{V}_{k,l}^* = \mathbf{V}_{k,l}(0)$ and $\xi_l=0$. Otherwise, we have
\begin{equation}\label{eq:comp_slackness2}
    \sum_{k \in \mathcal{K}_l} \tr \left( \mathbf{V}_{k,l} \mathbf{V}_{k,l}^\textrm{H} \right) = P^{\textrm{max}}.
\end{equation}
Let $\mathbf{D}_l \mathbf{\Lambda}_l \mathbf{D}_l^\textrm{H}$ denote the eigendecomposition of $\sum_{i \in \mathcal{K}}\omega_i \mathbf{H}_{i,l}^\textrm{H} \mathbf{X}_i \mathbf{H}_{i,l}$, where $\mathbf{D}_l \in \mathbb{C}^{N_\textrm{t} \times N_\textrm{t}}$ is a unitary matrix of eigenvectors and $\mathbf{\Lambda}_l \in \mathbb{C}^{N_\textrm{t} \times N_\textrm{t}}$ is a diagonal matrix of the corresponding eigenvalues. Equation \eqref{eq:comp_slackness2} can be equivalently expressed as
\begin{equation}\label{eq:comp_slackness3}
    \tr \left( \left( \mathbf{\Lambda}_l + \xi_l \mathbf{I}_{N_\textrm{t}} \right)^{-2} \mathbf{\Phi}_{l} \right) = P^{\textrm{max}},
\end{equation}
where $\mathbf{\Phi}_{l} \in \mathbb{C}^{N_\textrm{t} \times N_\textrm{t}}$ is defined as
\begin{equation}
    \begin{split}
        \mathbf{\Phi}_{l} \triangleq \mathbf{D}_l^\textrm{H} \sum_{k \in \mathcal{K}_l} \left(\omega_k\mathbf{H}_{k,l}^\textrm{H} \mathbf{Y}_k^\textrm{H} - \sum_{i \in \mathcal{K}} \omega_i \mathbf{H}_{i,l}^\textrm{H} \mathbf{X}_i^\textrm{H} \mathbf{Z}_{i,k,l}\right)\\
        \left(\omega_k\mathbf{H}_{k,l}^\textrm{H} \mathbf{Y}_k^\textrm{H} - \sum_{i \in \mathcal{K}} \omega_i \mathbf{H}_{i,l}^\textrm{H} \mathbf{X}_i^\textrm{H} \mathbf{Z}_{i,k,l}\right)^\textrm{H}\mathbf{D}_l.
    \end{split}
\end{equation}
Since $\mathbf{\Lambda}_l$ is a diagonal matrix, equation \eqref{eq:comp_slackness3} is equivalent to
\begin{equation}\label{eq:comp_slackness4}
    \sum_{n=1}^{N_\textrm{t}} \frac{\left[ \mathbf{\Phi}_l \right]_{n,n}}{\left( \left[ \mathbf{\Lambda}_l \right]_{n,n} + \xi_l \right)^2} = P^{\textrm{max}}.
\end{equation}
Instead of using iterative methods such as bisection search to solve \eqref{eq:comp_slackness4} for $\xi_l$ which incurs a high computational overhead, we propose using a DNN to directly approximate the solution. Specifically, the DNN is trained to learn the mapping from the input parameters $\left[\diag(\mathbf{\Phi}_l)^\top,\, \diag(\mathbf{\Lambda}_l)^\top,\, P^{\textrm{max}} \right]^\top$ to the scalar output $\xi_l$ that satisfies \eqref{eq:comp_slackness4}. After training, the DNN provides near-instantaneous inference and significantly reduces the runtime compared to iterative solvers. The training data for the DNN can be generated offline by solving \eqref{eq:comp_slackness4} using bisection search for a wide range of input values.
Finally, $\mu_k$ can be updated in each iteration by using gradient ascent as
\begin{equation}\label{eq:mu_update}
    \mu_k \xleftarrow{} \left[\mu_k + \delta_k \left( R_{k}^{\textrm{min}} - r_{k} \right)\right]^+, \quad k \in \mathcal{K},
\end{equation}
where $\delta_k$ is the step size for user $k$. 

Note that the calculation of the terms $\sum_{i \in \mathcal{K}}\omega_i \mathbf{H}_{i,l}^\textrm{H} \mathbf{X}_i \mathbf{H}_{i,l}$ and $\sum_{i \in \mathcal{K}} \omega_i \mathbf{H}_{i,l}^\textrm{H} \mathbf{X}_i^\textrm{H} \mathbf{Z}_{i,k,l}$ in \eqref{eq:V_WMMSE} requires access to the channel matrices from each O-RU $l$ to all users $k \in \mathcal{K}$. As $K$ increases, this requirement becomes prohibitive. To improve scalability, we only consider the channel matrices of users $k \in \mathcal{K}_l$ when determining the precoding matrices for O-RU $l$. Accordingly, the precoding matrix can be expressed as
\begin{align}\label{eq:V_WMMSE_simp}
        \widetilde{\mathbf{V}}_{k,l} = &\left(\sum_{i \in \mathcal{K}_l}\omega_i \mathbf{H}_{i,l}^\textrm{H} \mathbf{X}_i \mathbf{H}_{i,l} + \xi_l \mathbf{I}_{N_\textrm{t}}\right)^{-1} \nonumber\\ &\left(\omega_k\mathbf{H}_{k,l}^\textrm{H} \mathbf{Y}_k^\textrm{H} - \sum_{i \in \mathcal{K}_l} \omega_i \mathbf{H}_{i,l}^\textrm{H} \mathbf{X}_i^\textrm{H} \mathbf{Z}_{i,k,l}\right).
\end{align}
Recall that each user in a cell-free massive MIMO system is only served by a subset of O-RUs. Therefore, it is reasonable to restrict the computation at O-RU $l$ to users $k \in \mathcal{K}_l$.

In the proposed solution, matrices $\mathbf{W}$, $\mathbf{U}$, and $\mathbf{V}$ are iteratively updated until the convergence criterion is satisfied. However, such an algorithm may not be suitable for RT precoding as it may require many iterations to converge and involves complex operations such as multiple matrix inversions. To bypass this iterative process and reduce the computational complexity, in the next section we will leverage the structure of the iterative optimization algorithm as expert knowledge and propose a DRL-based framework for efficient precoding.
\section{Distributed Precoding in O-RAN} \label{sec:algorithm}

One straightforward data-driven precoding method is to use DNNs to directly determine the precoding matrices \cite{MIMO:precoding_DRL3, MIMO:power1, MIMO:precoding_DRL4}. However, as highlighted in \cite{MIMO:E2E1}, this direct approach may result in suboptimal sum-rate performance. The reason is that although DNNs can learn to allocate the transmit power, they may not be effective to mitigate inter-user interference, especially in high signal-to-noise ratio (SNR) scenarios with densely deployed users and O-RUs.
Moreover, this approach requires the DNNs to directly determine $\sum_{l \in \mathcal{L}} \left| \mathcal{K}_l \right| N_\textrm{t} N_\textrm{s}$ complex values, which results in large input and output spaces and scalability challenges.

\begin{rev}
To address this issue, we leverage the iterative algorithm in Section \ref{subsec:optimal_structure} as communication-domain expert knowledge and propose a multi-agent DRL algorithm to determine the receive filter and weight matrices, $\mathbf{U}_k$ and $\mathbf{W}_k$, once per near-RT loop. The precoding matrices are then determined in each RT loop using the closed-form update equation \eqref{eq:V_WMMSE_simp}. This hybrid approach has two advantages. First, it bypasses the iterative procedure of the optimization algorithm. The DRL algorithm learns to determine $\mathbf{U}_k$ and $\mathbf{W}_k$ in a single forward pass. Thus, the precoding matrices can be directly obtained via \eqref{eq:V_WMMSE_simp} in a single iteration. Second, since the DRL algorithm only determines the low-dimensional variables $\mathbf{U}_k$ and $\mathbf{W}_k$, the action space dimension is reduced to $2KN_\textrm{r} N_\textrm{s} + 2KN_\textrm{s}^2$, which improves training stability, scalability, and convergence. This approach will be elaborated in the following subsections.
\end{rev}

\begin{rev}
    Note that, according to \eqref{eq:W_WMMSE} and \eqref{eq:U_WMMSE}, the optimal matrices $\mathbf{W}_k^*$ and $\mathbf{U}_k^*$ depend only on $\mathbf{\Xi}_{k,i} = \sum_{l \in \mathcal{L}^\textrm{UE}_i} \mathbf{H}_{k,l} \mathbf{V}_{i,l}, \; i \in \mathcal{K}$, from \eqref{eq:effective_matrix}, which represent the sum of the effective channels from multiple serving O-RUs. This summation leads to a degree of channel hardening, and $\mathbf{\Xi}_{k,i}$ exhibits reduced sensitivity to instantaneous small-scale fading variations. This effect becomes more significant as the number of antennas $N_\mathrm{t}$ per O-RU and the number of serving O-RUs $L_k^\textrm{UE}$ increase. Although channel hardening is generally weaker in cell-free massive MIMO compared to cellular massive MIMO systems where the antennas are co-located, it can still reduce the channel variability \cite{CF-MMIMO, Bjornson:CF-mMIMO}. Consequently, $\mathbf{U}_k$ and $\mathbf{W}_k$ vary more smoothly over time and are less sensitive to rapid fluctuations in the channel.
    On the other hand, the precoding matrices $\mathbf{V}_{k,l}$ depend directly on the instantaneous channel matrices $\mathbf{H}_{k,l}$, as shown in \eqref{eq:V_WMMSE}, and are more sensitive to channel variations. Hence, updating $\mathbf{U}_k$ and $\mathbf{W}_k$ at the near-RT timescale provides a reasonable tradeoff between accuracy and computational complexity, whereas the precoding matrices are updated at the RT timescale due to their dependence on the instantaneous channel realizations.
\end{rev}

\subsection{Obtaining $\mathbf{U}$ and $\mathbf{W}$ via Multi-Agent DRL}\label{subsec:MADRL}

Most recent works utilizing expert knowledge have adopted a centralized approach, where a single DNN takes the channel matrices of all users as input to determine the intermediate variables \cite{MIMO:E2E1,MIMO:E2E2, MIMO:WMMSE-unfolding}. Such an approach faces scalability challenges in cell-free massive MIMO systems with dense O-RU deployment, where each user is served by multiple O-RUs. To resolve this issue, we define a Markov game, where a DRL agent is assigned to each user $k \in \mathcal{K}$ to locally determine $\mathbf{U}_k$ and $\mathbf{W}_k$ for that user.
A Markov game is a mathematical framework that extends the Markov decision processes (MDPs) to multi-agent systems. It models an environment where multiple agents act sequentially according to their own observations and policies. The state of the environment evolves according to the joint action of the agents.
Each agent $k$ has an observation space $\mathcal{O}_k$, an action space $\mathcal{A}_k$, a policy $\pi_k$, and a reward function $R_k$. The environment has a state space $\mathcal{S}$. 
Each near-RT loop is treated as one step of the Markov game. Each non-RT loop corresponds to an episode. 
At each step $n$, agent $k$ receives an observation $o_k(n) \in \mathcal{O}_k$ from the state $s(n) \in \mathcal{S}$ via its observation function $O_k: \mathcal{S} \to \mathcal{O}_k$. The agent then selects an action $a_k(n) \in \mathcal{A}_k$ according to its policy $\pi_k: \mathcal{O}_k \to \mathcal{A}_k$. 
Given the state $s(n)$ and the joint action $\mathbf{a}(n)=\left(a_1(n),\ldots,a_K(n)\right) \in \mathcal{A}_1 \times \ldots \times \mathcal{A}_K$, agent $k$ receives a reward $R_k(n) \in \mathbb{R}$ according to its reward function $R_k:\mathcal{S}\times \mathcal{A}_1 \times \ldots \times \mathcal{A}_K \to \mathbb{R}$. The environment then transitions to the next state $s(n+1)$ according to the stochastic transition function $p:\mathcal{S} \times \mathcal{A}_1 \times \ldots \times \mathcal{A}_K \to \Delta(\mathcal{S})$, where $\Delta(\mathcal{S})$ denotes the set of probability distributions over $\mathcal{S}$.

\begin{rev}
The observation $o_k(n) \in \mathcal{O}_k$ of agent $k$ at step $n$ should contain the information necessary to determine the optimal matrices $\mathbf{U}_k^*$ and $\mathbf{W}_k^*$ for user $k$. 
From \eqref{eq:W_WMMSE} and \eqref{eq:U_WMMSE}, we notice that $\mathbf{U}_k^*$ and $\mathbf{W}_k^*$ are functions of $\mathbf{\Xi}_{k,i},\; i \in \mathcal{K}$, and can be expressed as $\mathbf{U}_k^* = f\left(\left\{\mathbf{\Xi}_{k,i}\right\}_{i \in \mathcal{K}}\right)$ and $\mathbf{W}_k^* = h\left(\left\{\mathbf{\Xi}_{k,i}\right\}_{i \in \mathcal{K}}\right)$. Thus, by providing matrices $\mathbf{\Xi}_{k,i},\; i \in \mathcal{K}$, as observation to agent $k$, it can learn the functions $f(\cdot)$ and $h(\cdot)$.
\end{rev}
However, including all $\mathbf{\Xi}_{k,i},\; i \in \mathcal{K}$, in observation space $\mathcal{O}_k$ can lead to a large state space and convergence issues in dense deployments with many users. To address this issue, we define $\mathcal{I}_k$, which is the set of $I$ users that have the most similar large-scale fading profiles to user $k$. The similarity between users $k$ and $i$ is quantified by the score $\sum_{l \in \mathcal{L}}\beta_{i,l}\beta_{k,l}$. In other words, set $\mathcal{I}_k$ includes user $k$ as an element as well as those $I-1$ users that can cause the strongest interference to user $k$. Consequently, the observation of agent $k$ at step $n$ is restricted to $\mathbf{\Xi}_{k,i},\; i \in \mathcal{I}_k$ from the most recent RT loop, i.e., $o_k(n)= \left[\left[\mathbf{\Xi}_{k,i}(nN_\textrm{RT} - 1)\right]_{i \in \mathcal{I}_k}\right]$. These matrices are provided to the near-RT RIC over the E2 interface at each near-RT loop.

In the proposed multi-agent DRL framework, each agent $k$ updates the matrices $\mathbf{U}_k$ and $\mathbf{W}_k$ for user $k$ once per near-RT loop. Thus, the action space consists of the matrices $\mathbf{U}_k$ and $\mathbf{W}_k$ and has a dimension of $2N_\textrm{r}N_\textrm{s} + 2N_\textrm{s}^2$. As a comparison, in the direct precoding approach, each agent $k$ directly determines the precoding matrices for user $k$, which results in an action space of dimension $2 N_\textrm{t} N_\textrm{s} L^\textrm{UE}_k$. 
\begin{rev}
We use the update equations from the iterative algorithm in Section \ref{subsec:optimal_structure} as expert knowledge to further reduce the size of the action space. Specifically, according to \eqref{eq:MSE} and \eqref{eq:W_WMMSE}, the optimal matrix $\mathbf{W}_k^*$ is positive definite. To enforce this structure, we construct $\mathbf{W}_k$ via its Cholesky decomposition as
\begin{equation}\label{eq:cholesky}
    \mathbf{W}_k = \mathbf{L}_k\mathbf{L}_k^\textrm{H},
\end{equation}
where $\mathbf{L}_k$ is a lower-triangular matrix with strictly positive real diagonal elements. This approach has two benefits. First, it ensures $\mathbf{W}_k$ is always positive definite. Second, it reduces the number of real-valued parameters to be learned. Determining the complex matrix $\mathbf{W}_k \in \mathbb{C}^{N_\textrm{s}\times N_\textrm{s}}$ requires $2N_\textrm{s}^2$ real parameters to be learned. On the other hand, the Cholesky factor $\mathbf{L}_k$ has $N_\textrm{s}$ real diagonal elements and $\frac{N_\textrm{s}(N_\textrm{s}-1)}{2}$ complex lower-triangular off-diagonal elements, which results in a total of $N_\textrm{s}^2$ real parameters.
Consequently, the action space dimension is reduced to $2N_\textrm{r}N_\textrm{s} + N_\textrm{s}^2$. The action $a_k(n) \in \mathcal{A}_k$ chosen by agent $k$ at step $n$ can be expressed as $a_k(n) = \left[\tril \left(\mathbf{L}_k(n)\right), \mathbf{U}_k(n)\right]$.
\end{rev}

The reward function of agent $k$ at step $n$ is defined as $R_k(n)=\frac{1}{N_\textrm{RT}}\sum_{t=nN_\textrm{RT}}^{nN_\textrm{RT} + N_\textrm{RT} -1}r_k(t)$, i.e., the average throughput of user $k$ during near-RT loop $n$. At each step $n$, agent $k$ selects an action that maximizes its expected discounted return $G_k(n) = \sum_{i=n}^{N_\textrm{nRT} - 1} \gamma^{i-n} R_k(i)$ throughout the rest of the episode, where $\gamma$ is a discount factor.

To learn an optimal policy, each agent $k$ must explore and evaluate the quality of different actions in a given state. The action-value function $Q_k(s,\mathbf{a})$ represents the expected discounted return for agent $k$ when the system starts in state $s$, the agents choose the joint action $\mathbf{a}$, and then follow the joint policy $\boldsymbol{\pi}=\left(\pi_1,\ldots, \pi_K\right)$. It is given by
\begin{equation}
    Q_k(s, \mathbf{a}) = \mathbb{E}_{\boldsymbol{\pi}} \left[G_k(n) \; | \; s(n)=s,\;\mathbf{a}(n)=\mathbf{a}\right].
\end{equation}

In this paper, we propose a multi-agent extension of the SAC algorithm introduced in \cite{ML:SAC} to deploy and train DRL agents. SAC is an off-policy actor-critic algorithm that maximizes a combination of the reward and policy entropy to encourage exploration. The overall objective function of agent $k$ is
\begin{equation}\label{eq:actor_obj}
    J(\pi_k) = \mathbb{E}_{\boldsymbol{\tau} \sim \boldsymbol{\pi}}\left[\sum_{n=0}^{N_\textrm{nRT} - 1} \gamma^n\left(R_k(n) + \alpha_k \mathcal{H}(\pi_k(\cdot | o_k(n)))\right)\right],
\end{equation}
where $\boldsymbol{\tau}=\{\mathbf{o}(n),\mathbf{R}(n)\}_{n=0}^{N_\textrm{nRT} - 1}$ is a trajectory. $\mathbf{o}(n)=\left(o_1(n),\ldots,o_K(n)\right)$ and $\mathbf{R}(n)=\left(R_1(n),\ldots,R_K(n)\right)$ denote the joint observations and rewards, respectively. A trajectory is generated by starting from a random state $s(0)$ and following the joint policy $\boldsymbol{\pi}$ until the end of the episode.
The entropy term $\mathcal{H}(\pi_k(\cdot | o_k(n))) = -\mathbb{E}_{a_k \sim \pi_k(\cdot | o_k)}[\log \pi_k (a_k(n) | o_k(n))]$ measures the uncertainty of policy $\pi_k$. 
By maximizing the objective function \eqref{eq:actor_obj}, SAC aims to maximize the reward of agent $k$ while also keeping the policy $\pi_k$ stochastic so that agent $k$ can explore new actions and avoid premature convergence to suboptimal policies.
$\alpha_k$ is the temperature parameter that balances exploration and exploitation. 
$\pi_k$ is referred to as the actor for agent $k$, which is approximated using a DNN with parameters $\theta^\pi_k$ and deployed as an xApp at the near-RT RIC. The actor objective function of agent $k$ is
\begin{multline}\label{eq:actor_loss}
    J_{\pi_k}(\theta^\pi_k) = \mathbb{E}_{\mathbf{o} \sim \mathcal{B}} \bigr[\mathbb{E}_{\mathbf{a} \sim \boldsymbol{\pi}(\cdot | \mathbf{o})}[\alpha_k \log \pi_k(a_k|o_k) - \\  \min_{i=1,2}Q_{k, i}(\mathbf{o}, \mathbf{a})]\bigr],
\end{multline}
where $\mathcal{B}$ is the experience replay buffer that contains the tuples $\left(\mathbf{o},\mathbf{o}', \mathbf{a}, \mathbf{R}\right)$, recording the experiences of all agents throughout training. $\mathbf{o}'$ denotes the joint observations after taking the joint action $\mathbf{a}$ and transitioning to the next state. To mitigate overestimation bias, SAC trains two separate Q-functions $Q_{k,1}$ and $Q_{k,2}$ with independent parameters $\theta_{k,1}^Q$ and $\theta_{k,2}^Q$, respectively. Throughout training, SAC uses the minimum of the two Q-values $\min_{i=1,2}Q_{k, i}(\mathbf{o}, \mathbf{a})$ to obtain a less biased estimate of the Q-value. This is called the critic for agent $k$, deployed as an rApp at the non-RT RIC. To stabilize training and avoid selfish policies, the critic for each agent $k$ has global awareness. It takes the collective states and actions of all agents as input and outputs the Q-value for agent $k$. The critic loss function of agent $k$ is defined as
\begin{algorithm}[t]
    \caption{Training procedure for the proposed multi-agent DRL algorithm} \label{alg:DRL_training}
    Initialize parameters $\theta^\pi_k$, $\theta_{k,1}^{Q}$, $\theta_{k,2}^{Q}$, $\hat{\theta}_{k,1}^Q$, and $\hat{\theta}_{k,2}^Q$ for each agent $k$\\
    \For{\textnormal{iteration} $:= 1$ \textnormal{to} $M_\textnormal{iter}$}{
        Set $m_\textnormal{frames}:=0$\\
        \While{$m_\textnormal{frames} \leq M_\textnormal{frames}$}{
        Observe initial state $\mathbf{s}(0)$\\
        \For{$n:=0$ \textnormal{to} $N_\textnormal{nRT} - 1$}{
        For each agent $k$, select action $a_k(n) := \pi_k\left(s_k(n)\right)$\\
        Execute actions $\mathbf{a}(n)$ and observe reward $\mathbf{R}(n)$ and new states $\mathbf{s}(n+1)$\\
        Store $\left(\mathbf{s}(n),\mathbf{s}(n+1), \mathbf{a}(n), \mathbf{R}(n)\right)$ in $\mathcal{B}$\\
        $m_\textnormal{frames} := m_\textnormal{frames} + 1$
        }
        }
        \For{\textnormal{optimizer\_step} $:= 1$ \textnormal{to} $M_\textnormal{opt}$}{
        Sample a batch of $M_\textnormal{batch}$ samples from $\mathcal{B}$\\
        For each agent $k$, update the critic by minimizing the loss in \eqref{eq:critic_loss}\\
        For each agent $k$, update the actor by minimizing the loss in \eqref{eq:actor_loss}\\
        For each agent $k$, update the temperature parameter by minimizing the loss in \eqref{eq:alpha_loss}\\
        For each agent $k$, update the target network parameters using \eqref{eq:target_Q}\\
        }
    }
    \textbf{Output:} Actor parameters $\theta^\pi_k$
\end{algorithm}
\setlength{\textfloatsep}{8pt}
\begin{multline}\label{eq:critic_loss}
    J_{Q_k,i}(\theta_{k,i}^Q) = \mathbb{E}_{(\mathbf{o},\mathbf{o}',\mathbf{a},\mathbf{R}) \sim \mathcal{B}}\left[\frac{1}{2}\left(Q_{k,i}(\mathbf{o},\mathbf{a}) - y_k\right)^2\right],\\ i \in \{1,2\},
\end{multline}
\begin{multline}\label{eq:y}
    y_k = R_k + \gamma \mathbb{E}_{\mathbf{a}' \sim \boldsymbol{\pi}(\cdot|\mathbf{o}')}\biggr[\min_{i=1,2}\hat{Q}_{k, i}(\mathbf{o}', \mathbf{a}') \\
    -\; \alpha_k \log \pi_k(a_k' | o_k')\biggr],
\end{multline}
where $\hat{Q}_{k, 1}$ and $\hat{Q}_{k, 2}$ are the target Q-functions of agent $k$ parameterized by $\hat{\theta}_{k,1}^Q$ and $\hat{\theta}_{k,2}^Q$, respectively. These parameters are updated throughout training as
\begin{equation}\label{eq:target_Q}
    \hat{\theta}^Q_{k,i}(n+1) = \upsilon_\theta \theta^{Q}_{k,i}(n+1) + (1-\upsilon_\theta) \hat{\theta}^Q_{k,i}(n), \quad i \in \{1,2\},
\end{equation}
where $\upsilon_\theta$ is the soft update rate. The temperature parameter $\alpha_k$ is updated via stochastic gradient descent so that the policy $\pi_k$ maintains a desired entropy level $\bar{\mathcal{H}}$ throughout training:
\begin{multline}\label{eq:alpha_loss}
    \alpha_k(n+1) = \alpha_k(n) + \upsilon_\alpha \mathbb{E}_{o_k \sim \mathcal{B}} \bigr[\mathbb{E}_{a_k \sim \pi_k(\cdot | o_k)}\\ [\log \pi_k(a_k|o_k) + \bar{\mathcal{H}}]\bigr],
\end{multline}
where $\upsilon_\alpha$ is the temperature learning rate. During training, the critic estimates the Q-value of the actor’s actions, while the actor refines its policy by minimizing the loss in \eqref{eq:actor_loss}.
When the training is complete, the actors determine $\mathbf{U}$ and $\mathbf{W}$ using their learned policies. 
The training procedure for the proposed multi-agent DRL algorithm is summarized in Algorithm \ref{alg:DRL_training}.

\begin{rev}
In our proposed framework, each agent has its own local observation space, action space, and reward function. These components have the same structure and dimensions across all agents. Thus, we can use parameter sharing to improve the efficiency of the proposed framework. All agents share the same actor, critic, and target critic parameters, i.e., $\theta^\pi_k=\theta^\pi$, $\theta_{k,i}^{Q}=\theta_{i}^{Q}$, and $\hat{\theta}_{k,i}^Q=\hat{\theta}_{i}^Q$ for $k \in \mathcal{K}$ and $i \in \{1,2\}$. Parameter sharing provides several advantages. First, it improves scalability. Instead of deploying separate parameters for each agent, we deploy a shared set of parameters and use it for all agents. Second, it accelerates training and improves convergence since the number of trainable parameters is significantly reduced. Third, the shared parameters are updated using the aggregated experience of all agents with diverse channel conditions. This results in a richer training distribution and improved generalization. Fourth, new users that join the network after training can directly utilize the shared parameters without incurring additional training overhead. Note that parameter sharing does not result in identical actions across agents, since each agent has a different local observation and thus, a different input to the actor network.
\end{rev}

\begin{rev}
The proposed framework also supports user mobility, which leads to changes in user-O-RU associations. This is because the observation and action space dimensions of the DRL agents are independent of the specific user–O-RU associations. Specifically, the input to agent $k$ is $\mathbf{\Xi}_{k,i} = \sum_{l \in \mathcal{L}^\textrm{UE}_i} \mathbf{H}_{k,l} \mathbf{V}_{i,l}, \; i \in \mathcal{I}_k$, which represents the aggregated effective channels from the serving O-RUs. The structure and dimension of $\mathbf{\Xi}_{k,i}$ are independent of the set and number of serving O-RUs. Furthermore, the outputs $\mathbf{L}_k$ and $\mathbf{U}_k$ are defined per user. Therefore, changes in user-O-RU associations do not require any modifications to the architecture of the DRL agents.
\end{rev}

\subsection{Distributed Precoding}
The precoding matrices need to be updated in each RT loop. However, the control loops within the near-RT RIC operate on a near-RT timescale (10 ms to 1 sec) and are therefore not suitable for RT precoding. To overcome this limitation, the final precoding step is offloaded to the O-DUs. Specifically, each O-DU $u \in \mathcal{U}$ hosts two dApps. The first dApp uses the pretrained DNN to determine $\xi_l, \; l \in \mathcal{L}_u^\textrm{DU}$, in each RT loop. The second dApp sequentially determines the precoding matrices $\mathbf{V}_{k,l}, \;k \in \mathcal{K}_l, l \in \mathcal{L}_u^\textrm{DU}$, using \eqref{eq:V_WMMSE_simp} in each RT loop. 

Recall that in cell-free massive MIMO with coherent joint transmission, the O-RUs jointly serve users and their precoding matrices are coupled. Two O-RUs $l$ and $j$ are coupled if they serve at least one common user, i.e., $\mathcal{K}_l \cap \mathcal{K}_j \neq \emptyset$. According to \eqref{eq:C_term} and \eqref{eq:V_WMMSE_simp}, determining the term $\mathbf{Z}_{i,k,l}, \; i \in \mathcal{K}_l, k \in \mathcal{K}_l$, for O-RU $l$ requires access to the channel and precoding matrices of the O-RUs coupled with it, i.e., $\mathbf{H}_{i,j}$ and $\mathbf{V}_{k,j}$ for $j \in \mathcal{L}_k^\textrm{UE} \setminus \{l\}$. Suppose O-RU $l$ is connected to O-DU $u$, i.e., $l \in \mathcal{L}_u^\textrm{DU}$. If a coupled O-RU $j$ is also connected to the same O-DU, i.e., $j \in \mathcal{L}_u^\textrm{DU}$, then its most recent channel and precoding matrices are locally available at O-DU $u$. Otherwise, if a coupled O-RU $j$ is served by a different O-DU, then its channel and precoding matrices are not locally available at O-DU $u$. To address this issue, we define an inter-O-DU interface called the D2 interface that enables information exchange across O-DUs. In each near-RT loop, an O-DU $u$ receives the latest channel and precoding matrices that it needs from other O-DUs via a publish/subscribe mechanism.

\begin{figure}[t]
\center{\includegraphics[width=\linewidth]{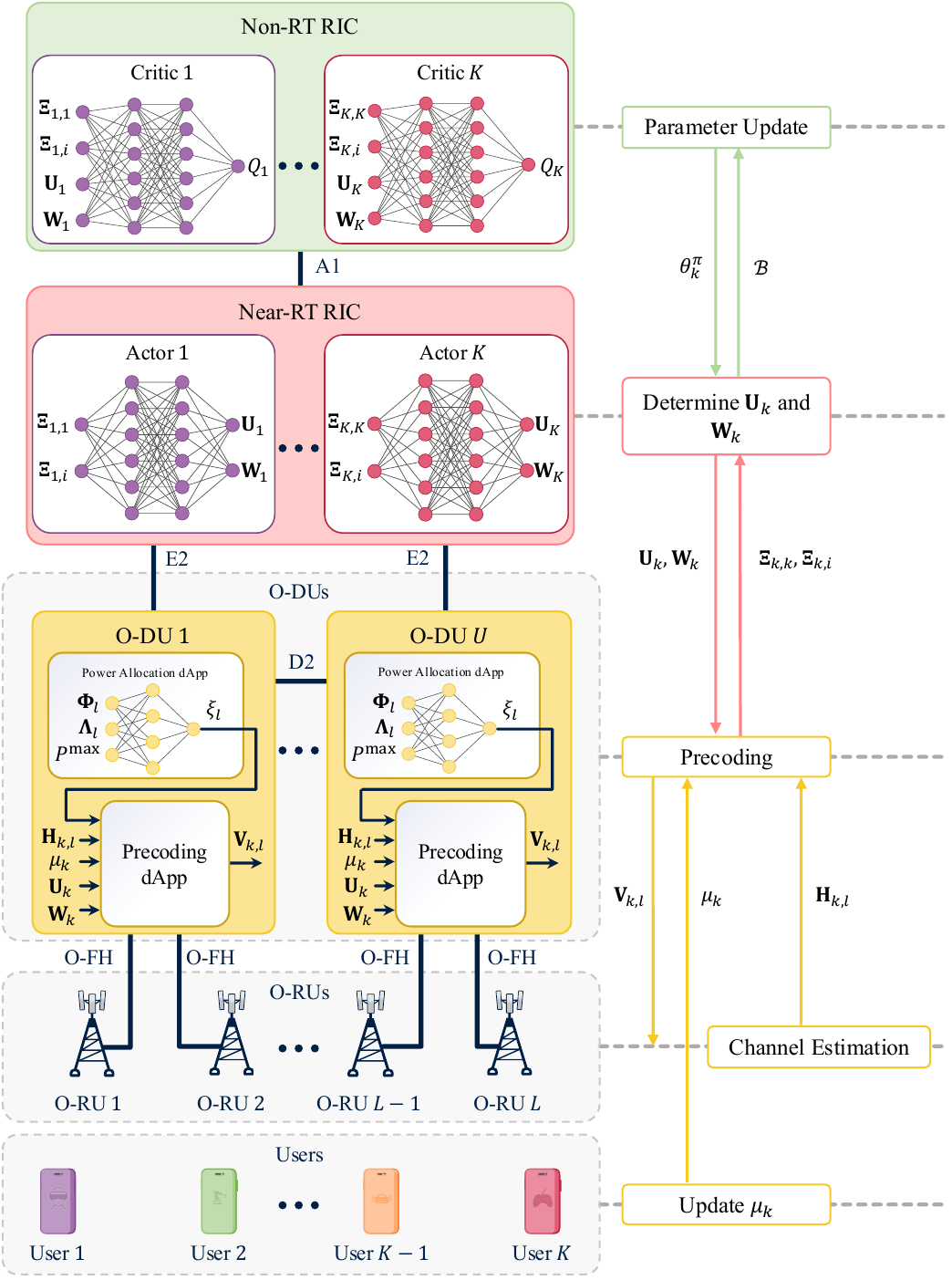}}
\caption{Block diagram of the proposed distributed precoding framework within O-RAN. The color of each block indicates its timescale. Green, red, and yellow blocks represent non-RT, near-RT, and RT operations, respectively.}
\label{fig:block_diagram}
\end{figure}

The stages of our hierarchical precoding framework are summarized as follows:

\subsubsection{RT Loop}
During each RT loop, each user $k \in \mathcal{K}$ updates $\mu_k$ using \eqref{eq:mu_update} according to the throughput it observed in the previous RT loop. It then sends the updated $\mu_k$ to its serving O-RUs $l \in \mathcal{L}^\textrm{UE}_k$ via uplink feedback. Each O-RU $l$ forwards this information along with the channel estimates $\mathbf{H}_{k,l}$ of users $k \in \mathcal{K}_l$ to its associated O-DU over the O-FH link. The O-DU $u$ then uses this information along with the latest $\mathbf{U}_k$ and $\mathbf{W}_k$ received from the near-RT RIC to determine the precoding matrices $\mathbf{V}_{k,l}, \;k \in \mathcal{K}_l,\; l \in \mathcal{L}_u^\textrm{DU}$, using \eqref{eq:V_WMMSE_simp}. This approach can dynamically adapt to changes in minimum data rate requirements. For instance, if the data rate requirement of user $k$ is increased in a given RT loop, $\mu_k$ will be updated accordingly via \eqref{eq:mu_update} during the next RT loop.

\subsubsection{Near-RT Loop}
In each near-RT loop, the near-RT RIC collects the most recent matrices $\mathbf{\Xi}_{k,k}$ and $\mathbf{\Xi}_{k,i}$ for $i\in \mathcal{I}_k, k \in \mathcal{K}$ from the O-DUs via the E2 termination. It stores them in the SDL database and notifies the xApps of the update using the internal messaging infrastructure. Then, each xApp $k$ queries the matrices $\mathbf{\Xi}_{k,k}$ and $\mathbf{\Xi}_{k,i}$ for $i\in \mathcal{I}_k$ from the database and obtains the updated $\mathbf{U}_k$ and $\mathbf{W}_k$. These matrices are subsequently sent to the O-DUs through the E2 termination. Note that $\mathbf{U}_k$ and $\mathbf{W}_k$ for user $k$ are sent to O-DU $u$ only if it serves at least one O-RU $l \in \mathcal{L}^\textrm{UE}_k$, i.e., if $\mathcal{L}^\textrm{DU}_u \cap \mathcal{L}^\textrm{UE}_k \neq \emptyset$. The near-RT RIC also collects the reward for the previous near-RT loop from the O-DUs to add a new experience to the replay buffer. Finally, O-DUs perform one round of information exchange with each other via the D2 interface in each near-RT loop.

\subsubsection{Non-RT Loop}
Each non-RT loop is treated as one episode in Algorithm \ref{alg:DRL_training}. It comprises $N_\textrm{nRT}$ near-RT loops. During each non-RT loop, a total of $N_\textrm{nRT}$ new experiences are added to the replay buffer. At the end of each non-RT loop, each rApp performs $M_\textrm{opt}$ optimization steps to update the actor and critic networks of the corresponding DRL agent.

Fig. \ref{fig:block_diagram} shows the block diagram of the proposed distributed precoding framework within O-RAN. Furthermore, the workflow of the proposed framework during a single non-RT loop is summarized in Algorithm \ref{alg:dist_precoding}.

\begin{algorithm}[t]
    \caption{The proposed distributed precoding algorithm for cell-free massive MIMO in O-RAN} \label{alg:dist_precoding}
    Initialize the precoding matrices $\mathbf{V}_{k,l}$ such that $\sum_{k \in \mathcal{K}_l} \tr\left(\mathbf{V}_{k,l}\mathbf{V}_{k,l}^\textrm{H}\right) \leq P^{\textrm{max}}, \; l \in \mathcal{L}$; Initialize $\xi_l:=0$. Initialize $\mu_k:=1$.\\
    \For{$n := 0$ \textnormal{to} $N_\textnormal{nRT} - 1$}{
        For each user $k \in \mathcal{K}$, sample $\mathbf{L}_k$ and $\mathbf{U}_k$ from the policy $\pi_k$ as $\left[\mathbf{L}_k, \mathbf{U}_k\right] \sim \pi_k \left(\cdot \mid \left[\left[\mathbf{\Xi}_{k,i}\right]_{i \in \mathcal{I}_k}\right]\right)$\\
        Determine $\mathbf{W}_k$ using \eqref{eq:cholesky}\\
        Send $\mathbf{U}_k$ and $\mathbf{W}_k$ to O-DU $u$ if $\mathcal{L}^\textrm{DU}_u \cap \mathcal{L}^\textrm{UE}_k \neq \emptyset$\\
        \For{$t = 0$ \textnormal{to} $N_\textnormal{RT} - 1$}{
            \For{\textnormal{each O-DU} $u \in \mathcal{U}$}{
                \For{\textnormal{each O-RU} $l \in \mathcal{L}^\textrm{DU}_u$}{
                    Update $\xi_l$ using \eqref{eq:comp_slackness4}\\
                    Update $\mathbf{V}_{k,l},\;k \in \mathcal{K}_l$ using \eqref{eq:V_WMMSE_simp}\\
                }
            }
            For each user $k \in \mathcal{K}$, update $\mu_k$ using \eqref{eq:mu_update}\\
        }
    }
    \textbf{Output:} Precoding matrices $\mathbf{V}_{k,l}$
\end{algorithm}
\setlength{\textfloatsep}{8pt}

\begin{rev}
\subsection{Computational Complexity Analysis}
In this subsection, we present the computational complexity of the proposed framework and discuss its scalability. The precoding matrices of each O-RU $l$ are updated according to \eqref{eq:V_WMMSE_simp} in each RT loop.
The dominant computational components arise from matrix multiplication and inversion operations.
Specifically, computing $\sum_{i \in \mathcal{K}_l}\omega_i \mathbf{H}_{i,l}^\textrm{H} \mathbf{X}_i \mathbf{H}_{i,l}$ has complexity $\mathcal{O} \left(K_l N_\textrm{r} N_\textrm{t}^2\right)$. The inversion of the resulting $N_\textrm{t} \times N_\textrm{t}$ matrix has complexity $\mathcal{O}\left(N_\textrm{t}^3\right)$. 
Furthermore, computing $\omega_k\mathbf{H}_{k,l}^\textrm{H} \mathbf{Y}_k^\textrm{H}$ for users $k \in \mathcal{K}_l$ has complexity $\mathcal{O}\left(K_l N_\textrm{r} N_\textrm{s} N_\textrm{t}\right)$. Finally, computing $\sum_{i \in \mathcal{K}_l} \omega_i \mathbf{H}_{i,l}^\textrm{H} \mathbf{X}_i^\textrm{H} \mathbf{Z}_{i,k,l} = \sum_{i \in \mathcal{K}_l} \omega_i \mathbf{H}_{i,l}^\textrm{H} \mathbf{X}_i^\textrm{H} \sum_{j \in \mathcal{L}^\textrm{UE}_k \setminus\{l\}} \mathbf{H}_{i,j} \mathbf{V}_{k,j}$ for users $k \in \mathcal{K}_l$ has complexity $\mathcal{O}\left(K_l^2 L^\textrm{UE} N_\textrm{r}^2 N_\textrm{t}\right)$. Thus, the overall computational complexity in each RT loop is $ \mathcal{O}\left(K_l^2 L^\textrm{UE} N_\textrm{r}^2 N_\textrm{t} + N_\textrm{t}^3\right)$. 

Note that the precoding matrices of each O-RU $l$ can be determined locally in each RT loop. Therefore, the computational complexity does not directly scale with the total number of users $K$ or total number of O-RUs $L$, which shows the scalability of the proposed framework.
\end{rev}

\section{Performance Evaluation} \label{sec:eval}
In this section, we evaluate the performance of our distributed precoding framework and compare it with different baselines.
In the considered cell-free O-RAN, the maximum transmit power of each O-RU is set to $P^{\textrm{max}} = 30$ dBm. The noise power is set to $\sigma^2= -114$ dBm. The number of antennas of each O-RU, $N_{\textrm{t}}$, is equal to 4 \cite{bjornson:O-RAN1}. The number of antennas of each user device, $N_{\textrm{r}}$, is equal to 2. The large-scale fading coefficients $\beta_{k,l}$ follow the 3GPP urban microcell non-line-of-sight (UMi-NLOS) pathloss model \cite{3GPP_pathloss} with a carrier frequency of $f_\textrm{c}=2 \textrm{ GHz}$. Considering a three-dimensional (3D) space with coordinates $[\text{x, y, z}]$, the z-coordinate of the O-RUs and users is fixed at $10$ and $2$, respectively.
For user-centric clustering, we select the $L_k^\textrm{UE}=L^\textrm{UE}=8$ O-RUs with the largest $\beta_{k,l}$ to serve user $k$. We set the observation cardinality $I=6$ for all DRL agents.
We set the same minimum data rate requirement of $R_k^\textrm{min}=R^\textrm{min}=4$ bits/s/Hz for all users $k \in \mathcal{K}$.
We set the user velocity $v_k$ to $1.4 \textrm{ m/s}$ ($5 \textrm{ km/hr}$).
We set the duration of each RT loop $T=1$ ms. A non-RT loop occurs once every $N_\textrm{nRT}=100$ near-RT loops, and each near-RT loop comprises $N_\textrm{RT}=10$ RT loops.
\begin{rev}
At the beginning of each non-RT loop, the user-O-RU associations are updated according to the new locations of the users and the updated values of $\beta_{k,l}$. Consequently, the agents are trained under dynamic user-O-RU associations, which prevents overfitting to a fixed deployment and enhances robustness to topology changes.
\end{rev}

We use the BenchMARL \cite{BenchMARL} and TorchRL \cite{TorchRL} libraries to implement the proposed algorithm. The discount factor $\gamma$ is set to $0.9$.
For Algorithm \ref{alg:DRL_training}, we set $M_\textnormal{iter} = 24000$, $M_\textnormal{frames} = 6000$, $M_\textnormal{opt} = 60$, and $M_\textnormal{batch} = 512$. The size of the replay buffer $\mathcal{B}$ is $10^5$. We set the soft update rate $\upsilon_\theta = 0.005$. 
For each DRL agent, the actor network is a DNN comprising two fully connected (FC) layers with 128 neurons each, while the critic network has two FC layers with 256 neurons each.
 
We consider the following baseline precoding schemes:
\begin{enumerate}
    \begin{rev}\item \textbf{Cell-free WMMSE (CF-WMMSE):} This baseline corresponds to the iterative optimization algorithm proposed in Section \ref{subsec:optimal_structure}. In this algorithm, the matrices $\mathbf{W}$, $\mathbf{U}$, and $\mathbf{V}$ are iteratively updated according to \eqref{eq:W_WMMSE}, \eqref{eq:U_WMMSE}, and \eqref{eq:V_WMMSE}, respectively, until convergence.\end{rev}
    \item \textbf{Centralized regularized zero-forcing (C-RZF):} C-RZF is a linear precoding scheme that eliminates inter-user interference by projecting each user’s signal onto the null space of all other users’ channels. C-RZF precoding requires global CSI and centralized processing. Let $\mathbf{H} \in \mathbb{C}^{K N_\textrm{r} \times L N_\textrm{t}}$ denote the concatenated channel matrix. The C-RZF precoding matrix is given by
    \begin{equation}
        \widetilde{\mathbf{V}}^\textrm{C-RZF} = \mathbf{H}^\textrm{H} \left( \mathbf{H} \mathbf{H}^\textrm{H} + \lambda \mathbf{I}_{KN_\textrm{r}}\right)^{-1},
    \end{equation}
    where $\lambda$ is the regularization parameter that improves robustness to noise and ill-conditioned channels. The above precoding matrix needs to be normalized to satisfy the power constraint. Similar to \cite{MIMO:clustering2}, we use a fractional power allocation method and define the normalized precoding matrix as $\mathbf{V}^\textrm{C-RZF} = \eta \widetilde{\mathbf{V}}^\textrm{C-RZF}$, where
    \begin{equation}
        \eta = \sqrt{\frac{P^\textrm{max}}{\max_{l \in \mathcal{L}}\left(\sum_{k \in \mathcal{K}_l} \tr\left(\widetilde{\mathbf{V}}^\textrm{C-RZF}_{k,l}\left(\widetilde{\mathbf{V}}^\textrm{C-RZF}_{k,l}\right)^\textrm{H}\right)\right)}}.
    \end{equation}

    \item \textbf{Distributed regularized zero-forcing (D-RZF):} D-RZF is a distributed variation of the original RZF precoding scheme, where each O-RU $l$ independently computes its precoding matrices using only local CSI of users $k \in \mathcal{K}_l$. Let $\mathbf{H}_l \in \mathbb{C}^{K_l N_\textrm{r} \times N_\textrm{t}}$ denote the concatenated channel matrix from O-RU $l$ to users $k \in \mathcal{K}_l$. The D-RZF precoding matrix at O-RU $l$ is given by
    \begin{equation}
        \widetilde{\mathbf{V}}_l^\textrm{D-RZF} =  \mathbf{H}_l^\textrm{H} \left( \mathbf{H}_l \mathbf{H}_l^\textrm{H} + \lambda \mathbf{I}_{K_lN_\textrm{r}}\right)^{-1}.
    \end{equation}
    Since in D-RZF each O-RU locally determines its precoding matrix, power normalization can also be performed independently at each O-RU as $\mathbf{V}_l^\textrm{D-RZF} = \eta_l\widetilde{\mathbf{V}}_l^\textrm{D-RZF}$, where $\eta_l$ is a normalization factor to satisfy the transmit power constraint at O-RU $l$, defined as
    \begin{equation}
        \eta_l = \sqrt{\frac{P^\textrm{max}}{\sum_{k \in \mathcal{K}_l} \tr\left(\widetilde{\mathbf{V}}^\textrm{D-RZF}_{k,l}\left(\widetilde{\mathbf{V}}^\textrm{D-RZF}_{k,l}\right)^\textrm{H}\right)}}.
    \end{equation}

    \begin{rev}
    \item \textbf{D-RZF with learning-based power allocation (D-RZF-LPA):} This approach has been proposed in \cite{D-RZF-LPA}. The precoding matrices $\widetilde{\mathbf{V}}_{k,l}^{\textrm{D-RZF-LPA}}$ are determined in the same manner as in D-RZF. A DNN is trained for each O-RU $l$ to determine the power allocation coefficients $\eta_{k,l}$ for each user $k \in \mathcal{K}_l$. The final precoding matrices are obtained as
    \begin{equation}
        \mathbf{V}_{k,l}^{\textrm{D-RZF-LPA}} = \eta_{k,l} \frac{\widetilde{\mathbf{V}}^\textrm{D-RZF-LPA}_{k,l}}{\sqrt{\tr\left(\widetilde{\mathbf{V}}^\textrm{D-RZF-LPA}_{k,l}\left(\widetilde{\mathbf{V}}^\textrm{D-RZF-LPA}_{k,l}\right)^\textrm{H}\right)}}.
    \end{equation}
    The input to DNN $l$ consists of the large-scale fading coefficients $\beta_{k,l}$ for each user $k \in \mathcal{K}_l$. The DNNs are trained in a supervised manner. The training dataset includes power allocation solutions obtained via an iterative optimization algorithm.
    \end{rev}
    
    \item \textbf{DRL-WMMSE:} This approach has been adopted in \cite{MIMO:precoding_DRL2}. In the original WMMSE algorithm proposed for cellular architectures, the precoding subproblem is decoupled across O-RUs. Each O-RU $l$ independently determines its precoding matrix for users $k \in \mathcal{K}_l$ as
    \begin{equation}\label{eq:V_WMMSE_original}
            \mathbf{V}_{k,l}^* = \left(\sum_{i \in \mathcal{K}}\omega_i \mathbf{H}_{i,l}^\textrm{H} \mathbf{X}_i \mathbf{H}_{i,l} + \xi_l \mathbf{I}_{N_\textrm{t}}\right)^{-1} \omega_k\mathbf{H}_{k,l}^\textrm{H} \mathbf{Y}_k^\textrm{H}.
    \end{equation}
    Similar to the proposed framework, we use the expert knowledge from the iterative algorithm and assign a DRL agent to each user $k$ to determine $\mathbf{U}_k$ and $\mathbf{W}_k$. The precoding matrices are determined using \eqref{eq:V_WMMSE_original}.
    \begin{rev}\item \textbf{End-to-end DRL (E2E-DRL):} In this approach, a DRL agent is assigned to each O-RU. Agent $l$ observes the local channel matrices $\mathbf{H}_{k,l}$ for each user $k \in \mathcal{K}_l$ and directly determines the corresponding precoding matrices $\mathbf{V}_{k,l},\; k \in \mathcal{K}_l$. Similar to D-RZF, the outputs of each agent $l$ are locally normalized to satisfy the transmit power constraint of O-RU $l$.  \end{rev}
\end{enumerate}

\begin{figure}[t]
\center{\includegraphics[height=50mm]{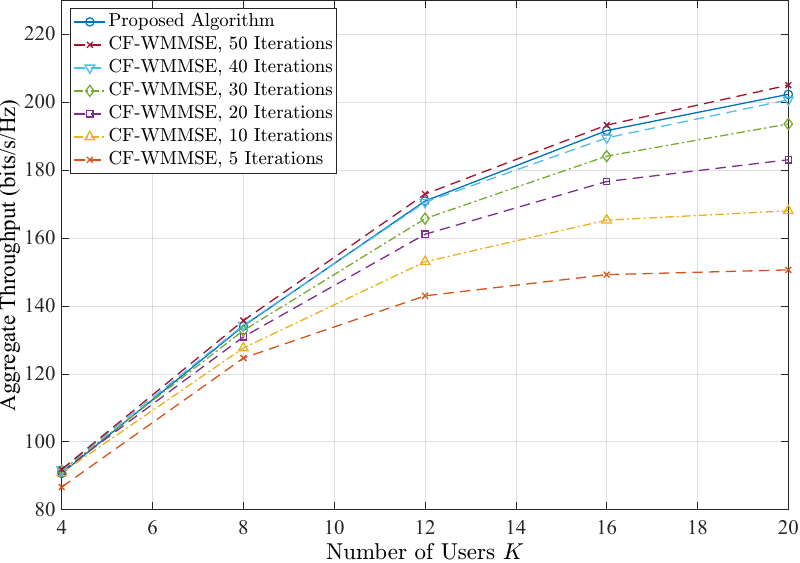}}
\caption{Aggregate throughput of the proposed algorithm and the CF-WMMSE algorithm with different numbers of iterations versus the number of users $K$ for $L=36$.}
\label{fig:K_WMMSE}
\end{figure}

\begin{rev}
First, in order to evaluate the effectiveness of the proposed expert-guided multi-agent DRL framework, we compare its performance with the CF-WMMSE algorithm. Note that the CF-WMMSE algorithm can be evaluated only in small-scale configurations with limited users and O-RUs due to its high computational complexity. Therefore, we first consider a small-scale setup with $L=36$ O-RUs randomly deployed in a $500 \textrm{ m} \times 500 \textrm{ m}$ area and served by a single O-DU.
Fig. \ref{fig:K_WMMSE} shows the aggregate throughput versus the number of users $K$ for the proposed framework and the CF-WMMSE algorithm with $5$, $10$, $20$, $30$, $40$, and $50$ iterations. For small numbers of users, only a few iterations are required for the convergence of the CF-WMMSE algorithm. Thus, the proposed framework and the CF-WMMSE algorithm with different numbers of iterations achieve similar performance. However, as $K$ increases, the performance gap between the CF-WMMSE algorithm with different numbers of iterations becomes larger, which indicates that more iterations are required for the algorithm to converge. The proposed framework achieves performance similar to CF-WMMSE with $50$ iterations, which shows the effectiveness of the proposed multi-agent DRL algorithm. 
\end{rev}

\begin{figure}[t!]
\center{\includegraphics[height=55mm]{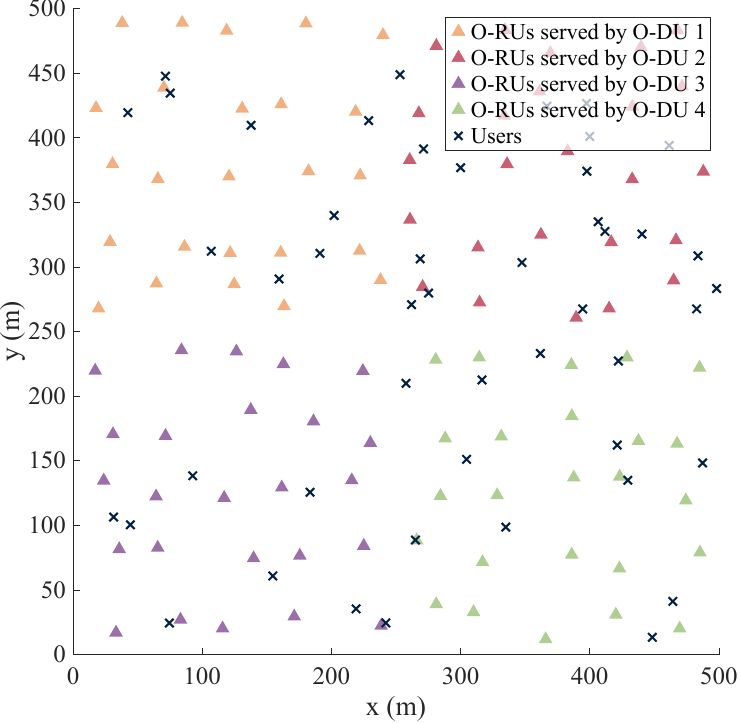}}
\caption{Topology of the considered cell-free O-RAN with $U=4$, $L=100$, and $K=48$. O-RUs served by the same O-DU are shown in the same color. Users are depicted at their initial locations.}
\label{fig:locations}
\end{figure}

In the remainder of the simulations, we consider a cell-free O-RAN consisting of $L=100$ O-RUs and $K=48$ users randomly deployed in a $500 \textrm{ m} \times 500 \textrm{ m}$ area. We divide the simulation area into $U=4$ square subareas and deploy one O-DU in each subarea to serve the O-RUs located within that subarea. We use a wrap-around topology to mimic a large network deployment. The considered cell-free O-RAN is illustrated in Fig. \ref{fig:locations}.

\begin{figure}[t!]
\center{\includegraphics[height=50mm]{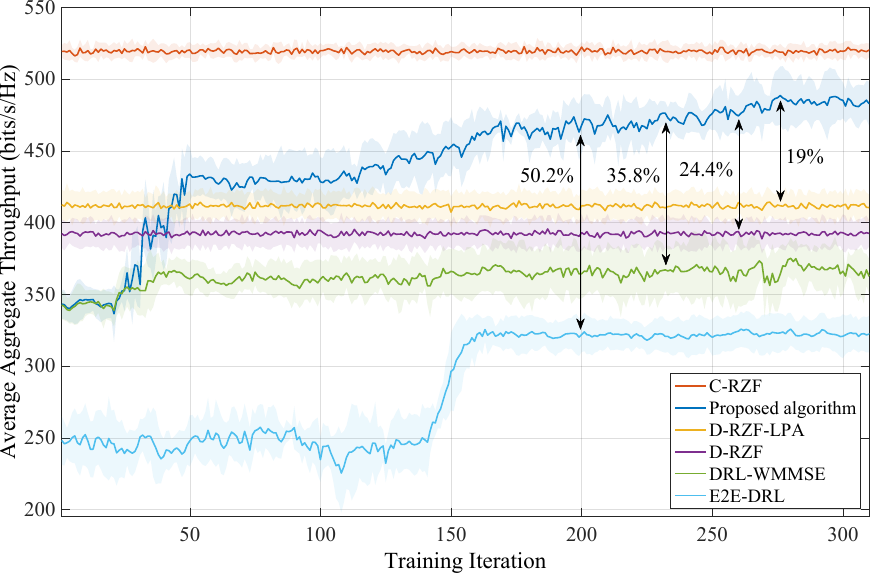}}
\caption{Convergence of the average aggregate throughput over non-RT loops during training. The shaded regions represent the standard deviation of the aggregate throughput.}
\label{fig:convergence}
\end{figure}

Fig. \ref{fig:convergence} shows the convergence of the average aggregate throughput for the proposed framework and the baselines over 10 random seeds. The shaded regions represent the standard deviation of the aggregate throughput at each training iteration. 
The proposed precoding scheme performs close to C-RZF due to the information exchange between O-DUs. 
It also outperforms the distributed precoding methods. 
In particular, it exceeds D-RZF-LPA and D-RZF by up to $19 \%$ and $24.4 \%$, respectively, as D-RZF relies solely on local CSI at each O-RU. 
Moreover, the proposed framework outperforms DRL-WMMSE by up to $35.75 \%$ since the solution obtained by the original WMMSE algorithm is suboptimal in cell-free massive MIMO with coherent joint transmission. 
\begin{rev}
The proposed framework also outperforms the E2E-DRL baseline by up to $50.2\%$. This is because the E2E-DRL approach attempts to learn the precoding matrices directly from high-dimensional channel observations, which leads to large state and action spaces, especially in dense deployments with many users and O-RUs. As a result, the agents struggle to learn effective interference-mitigation strategies. These results highlight the importance of incorporating expert knowledge into the learning process.
\end{rev}

\begin{figure}[t]
\centering
{\subfloat[]{\includegraphics[width=0.49\linewidth]{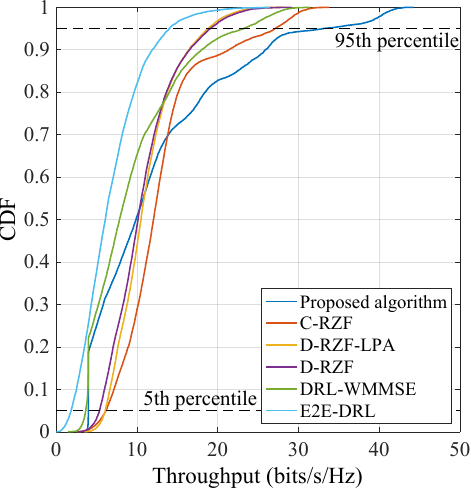}}}
{\subfloat[]{\includegraphics[width=0.49\linewidth]{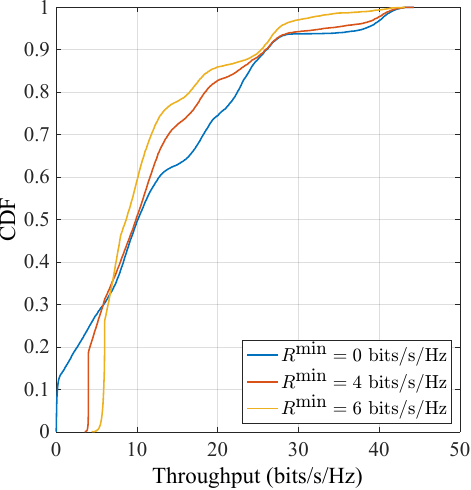}}}
\caption{CDF of per-user throughput for (a) the proposed algorithm and baselines (b) the proposed algorithm with $R^\textrm{min}=0$, $4$, and $6$ bits/s/Hz.}
\label{fig:CDF}
\end{figure}

Fig. \ref{fig:CDF}(a) shows the cumulative distribution function (CDF) of per-user throughput for the proposed framework and the baselines. For each throughput value on the x-axis, the CDF reflects the fraction of users whose throughput is less than or equal to that value. Examining the 5th and 95th percentiles reveals that the proposed framework yields a lower throughput for users with poor channel conditions but a higher throughput for users with favorable channel conditions when compared with RZF schemes. This is because the proposed algorithm is designed to maximize aggregate throughput. Thus, it prioritizes users with favorable channel conditions. However, unlike the original WMMSE algorithm in \cite{MIMO1}, which only focuses on maximizing the aggregate throughput, the fairness of the proposed framework can be controlled by adjusting $R^\textrm{min}$.

To evaluate the effect of $R^\textrm{min}$ on the fairness of the proposed framework, in Fig. \ref{fig:CDF}(b) we present the CDF curve of per-user throughput for different values of $R^\textrm{min}$. When $R^\textrm{min}$ is equal to $0$, the algorithm focuses only on maximizing the aggregate throughput, thus prioritizing users with favorable channel conditions. However, as we increase $R^\textrm{min}$, the algorithm sacrifices the data rates of high-throughput users to improve fairness and ensure that all users meet the minimum rate requirement. For example, when $R^\textrm{min}=4$ bits/s/Hz, the CDF curve has a shorter tail compared to the case with $R^\textrm{min}=0$, indicating that fewer users have very high data rates. However, all users achieve at least $r_k=4$ bits/s/Hz.

\begin{figure}[t!]
\center{\includegraphics[height=50mm]{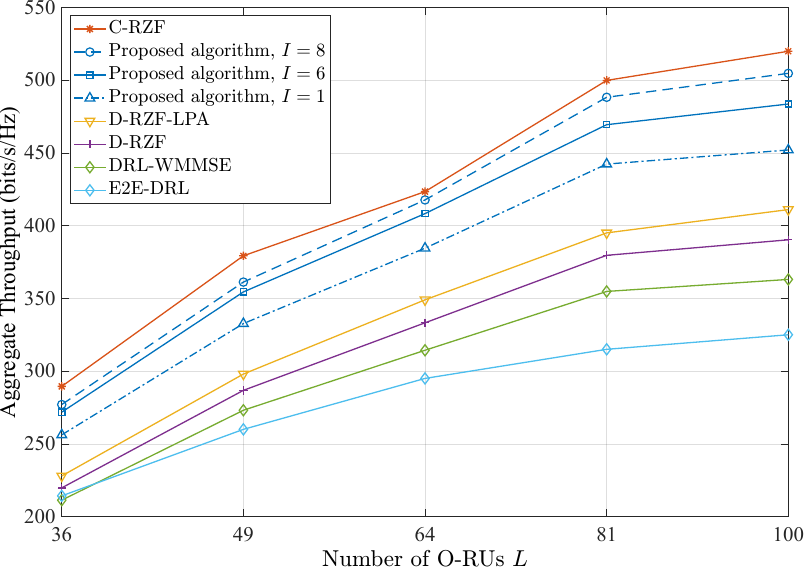}}
\caption{The aggregate throughput versus the number of O-RUs.}
\label{fig:ORUs}
\end{figure}


\begin{table*}[ht]
\begingroup
\centering
\caption{Comparison of precoding schemes in terms of computational complexity and training/inference latency}
\resizebox{\textwidth}{!}{%
\begin{tabular}{|l|c|c|c|c|c|c|}
\hline
Framework & C-RZF & Proposed algorithm & D-RZF & D-RZF-LPA & DRL-WMMSE & E2E-DRL \\
\hline
Precoding Complexity & 
\begin{tabular}[c]{@{}c@{}} 
\( \mathcal{O}\left(K^2 L N_\textrm{r}^2 N_\textrm{t} \right. \) \\ 
\( \left. + K^3 N_\textrm{r}^3\right) \)
\end{tabular} & 
\begin{tabular}[c]{@{}c@{}} 
\( \mathcal{O}\left(K_l^2 L^\textrm{UE} N_\textrm{r}^2 N_\textrm{t} \right. \) \\ 
\( \left. + N_\textrm{t}^3\right) \)
\end{tabular} & 
\begin{tabular}[c]{@{}c@{}} 
\( \mathcal{O}\left( K_l^2 N_\textrm{r}^2 N_\textrm{t} \right. \) \\ 
\( \left. + K_l^3 N_\textrm{r}^3 \right) \)
\end{tabular} & 
\begin{tabular}[c]{@{}c@{}} 
\( \mathcal{O}\left( K_l^2 N_\textrm{r}^2 N_\textrm{t} \right. \) \\ 
\( \left. + K_l^3 N_\textrm{r}^3 \right) \)
\end{tabular} & 
\begin{tabular}[c]{@{}c@{}} 
\( \mathcal{O}\left(K_l N_\textrm{r} N_\textrm{t}^2 \right. \) \\ 
\( \left. + N_\textrm{t}^3\right) \)
\end{tabular} & 
\( \mathcal{O}\left(K_l^2 N_\textrm{r}^2 N_\textrm{t}^2\right) \) \\
\hline
Training Iteration Duration (s) & -- & 64.88 & -- & 24.3 & 53.82 & 58.4 \\
\hline
Near-RT Loop Execution Time (ms) & 27.82 & 6.38 & 4.52 & 4.89 & 2.49 & 3.35 \\
\hline
\end{tabular}%
}
\label{tab:complexity_comparison}
\endgroup
\end{table*}

Fig. \ref{fig:ORUs} shows the aggregate throughput versus the number of O-RUs for the baseline schemes and for the proposed algorithm with the observation cardinality $I=1$, $6$, and $8$. In the case of $I=1$, each agent $k$ only observes its effective channel matrix $\mathbf{\Xi}_{k,k}$ without any interference information. The aggregate throughput increases with the number of O-RUs across all schemes. The proposed framework consistently outperforms the distributed baselines and performs close to C-RZF. Increasing $I$ improves performance but at the cost of higher input dimensionality and longer training time. Increasing $I$ from $1$ to $6$ yields a $6.38\%$ improvement on average, while increasing it from $6$ to $8$ adds only $2.9\%$ on average.

\begin{figure}[t]
\center{\includegraphics[height=50mm]{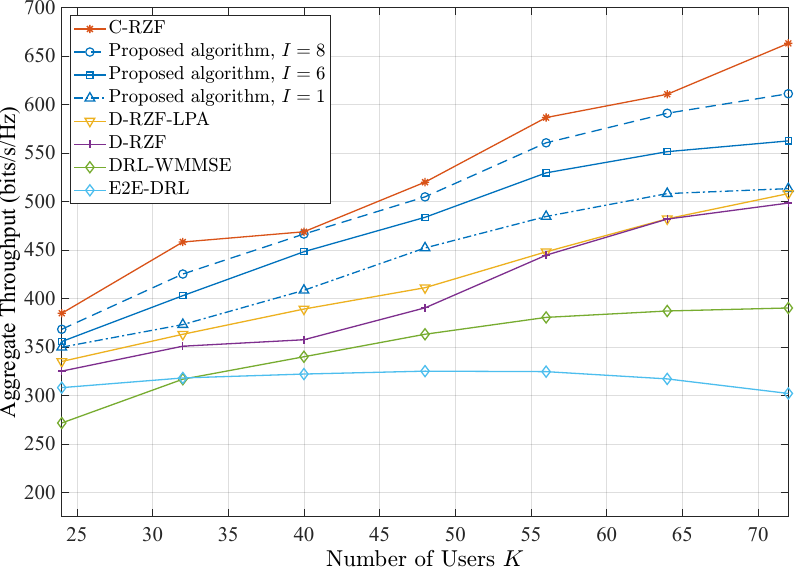}}
\caption{The aggregate throughput versus the number of users.}
\label{fig:K}
\end{figure}

Fig. \ref{fig:K} shows the aggregate throughput for varying numbers of users $K$. 
Consistent with the trend in Fig. \ref{fig:ORUs}, increasing the observation cardinality $I$ improves the performance of the proposed framework. As $K$ increases, the performance gap between the proposed algorithm with $I=1$, $6$, and $8$ increases. This is because a larger $K$ results in a denser user deployment and stronger inter-user interference. Thus, a larger observation cardinality is required to determine $\mathbf{U}^*$ and $\mathbf{W}^*$. 
\begin{rev}
It can also be observed that for $K=24$, E2E-DRL outperforms DRL-WMMSE and performs close to D-RZF. However, its performance deteriorates as $K$ increases. This further confirms that the E2E-DRL approach may struggle to learn effective interference-mitigation strategies when the number of users increases and inter-user interference becomes stronger.
\end{rev}

\begin{figure}[t]
\center{\includegraphics[height=45mm]{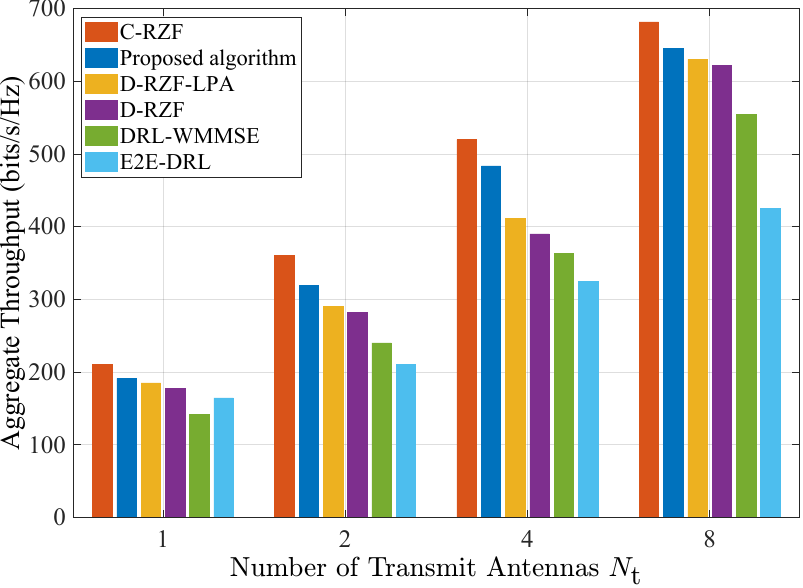}}
\caption{The aggregate throughput versus the number of transmit antennas.}
\label{fig:N_t}
\end{figure}

Fig. \ref{fig:N_t} shows the aggregate throughput versus the number of transmit antennas $N_\textrm{t}$ for the proposed framework and the baselines. It can be observed that the proposed framework consistently outperforms the distributed baselines and performs close to C-RZF across different values of $N_\textrm{t}$. Notably, as $N_\textrm{t}$ increases from $4$ to $8$, the performance gap between the proposed framework and D-RZF-LPA narrows from $19\%$ to $2.38\%$. This is because, beyond a certain point, each O-RU has sufficient spatial degrees of freedom to locally suppress interference and centralized precoding or inter-O-RU information exchange become less significant.

\begin{rev}
Table \ref{tab:complexity_comparison} presents a comparison of the computational complexity and the training and inference latency between the proposed framework and the baselines. The computational complexity of C-RZF scales cubically with the number of users $K$ and linearly with the number of O-RUs $L$. On the other hand, the computational complexities of the proposed algorithm, D-RZF-LPA, D-RZF, DRL-WMMSE, and E2E-DRL do not directly scale with $K$ and $L$ due to distributed precoding. The proposed framework incurs a higher training time compared with DRL-WMMSE and E2E-DRL. However, after training is completed, it requires only $6.38$ ms to execute a near-RT loop, which is well below both the $10$ ms target loop duration and the $27.82$ ms required by C-RZF.
\end{rev}

\begin{figure}[t]
\centering
{\subfloat[]{\includegraphics[height=43.5mm]{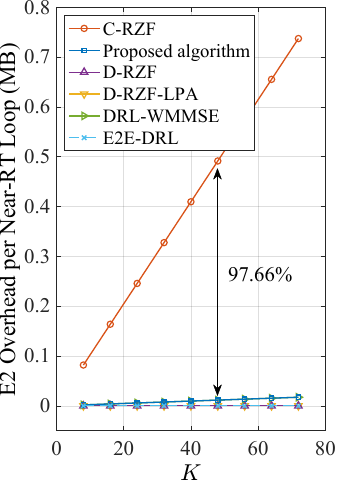}}}
{\subfloat[]{\includegraphics[height=43.5mm]{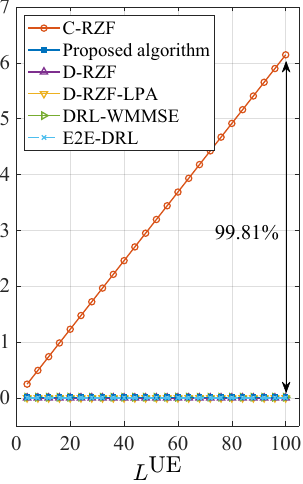}}}
{\subfloat[]{\includegraphics[height=43.5mm]{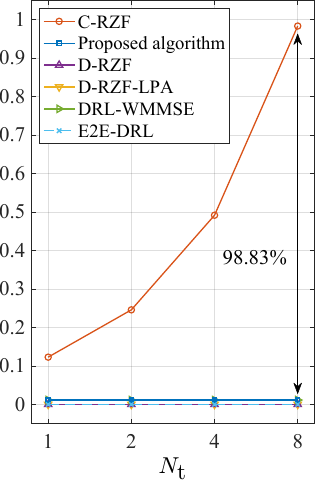}}}
\caption{E2 interface signaling overhead versus (a) number of users $K$, (b) cluster size $L^\textrm{UE}$, and (c) number of transmit antennas $N_\textrm{t}$.}
\label{fig:E2_overhead}
\end{figure}

\begin{rev}
Fig. \ref{fig:E2_overhead} compares the signaling overhead on the E2 interface between the proposed framework and the baselines versus the number of users $K$, cluster size $L^\textrm{UE}$, and number of transmit antennas $N_\textrm{t}$. In C-RZF, each O-DU $u$ transmits the channel matrices of O-RUs $l \in \mathcal{L}_u^\textrm{DU}$ to the near-RT RIC in each RT loop and receives the precoding matrices in return. On the other hand, the proposed framework sends only $\mathbf{\Xi}_{k,i}, \; i \in \mathcal{I}_k$, for each user $k$ and receives $\mathbf{U}_k$ and $\mathbf{W}_k$ once per near-RT loop. D-RZF-LPA, D-RZF, and E2E-DRL do not incur an overhead on E2 interface, since the precoding matrices are determined locally. DRL-WMMSE has the same overhead as the proposed framework. When the number of users $K$ increases, the signaling overhead of the proposed framework and C-RZF grows linearly. However, the proposed framework consistently reduces the overhead by $97.66\%$. Moreover, the signaling overhead of the proposed framework remains constant with respect to the cluster size $L^\textrm{UE}$ since only $\mathbf{\Xi}_{k,i}, \; i \in \mathcal{I}_k$, is transmitted instead of the raw channel matrices. Lastly, the overhead of the proposed framework is independent of the number of transmit antennas $N_\textrm{t}$ since $\mathbf{\Xi}_{k,i}$ and $\mathbf{U}_k$ are $N_\textrm{r} \times N_\textrm{s}$ matrices and $\mathbf{W}_k$ is an $N_\textrm{s} \times N_\textrm{s}$ matrix. When $N_\textrm{t} = 8$, the overhead is reduced by $98.83\%$ compared with C-RZF.
\end{rev}

\begin{figure}[t!]
\centering
{\subfloat[]{\includegraphics[height=45mm]{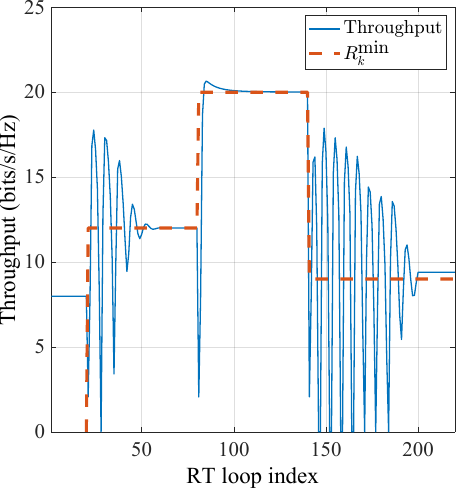}}}
{\subfloat[]{\includegraphics[height=45mm]{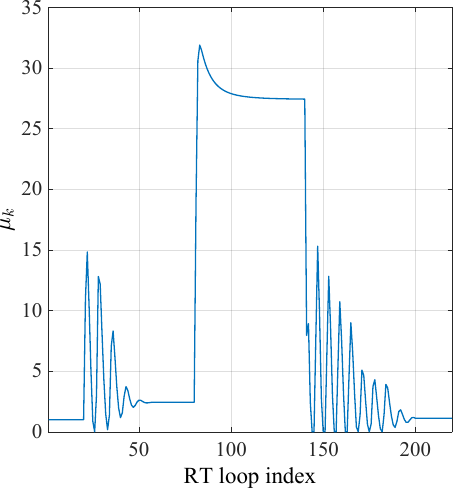}}}
\caption{(a) Throughput and (b) Lagrange multiplier of a random user $k$ with dynamic minimum data rate requirements over RT loop index after training.}
\label{fig:mu_convergence}
\end{figure}

Next, we vary the minimum data rate requirement of a randomly selected user after training to evaluate whether the proposed framework can adapt to such changes. Fig. \ref{fig:mu_convergence} shows the throughput $r_k$ and the Lagrange multiplier $\mu_k$ of this user after training. Initially, $R_k^\textrm{min} = 0$, and the throughput is stabilized at $r_k = 7.98$ bits/s/Hz with $\mu_k = 1$. At $t = 20$, the minimum data rate requirement is increased to $R_k^\textrm{min} = 12$ bits/s/Hz. Results in Fig. \ref{fig:mu_convergence} show that $\mu_k$ begins to oscillate and eventually stabilizes. The throughput $r_k$ converges to $12$ bits/s/Hz. Similar behavior is observed when the requirement is increased to $R_k^\textrm{min} = 20$ bits/s/Hz at $t = 80$ and then decreased to $R_k^\textrm{min} = 9$ bits/s/Hz at $t = 140$. In each case, the value of $\mu_k$ is updated quickly, and the throughput converges to the new $R_k^\textrm{min}$, confirming that the proposed framework can dynamically adapt to changes in minimum data rate requirements without fine-tuning.

\begin{figure}[t!]
\centering
{\subfloat[$N_\mathrm{t}=4$]{\includegraphics[height=45mm]{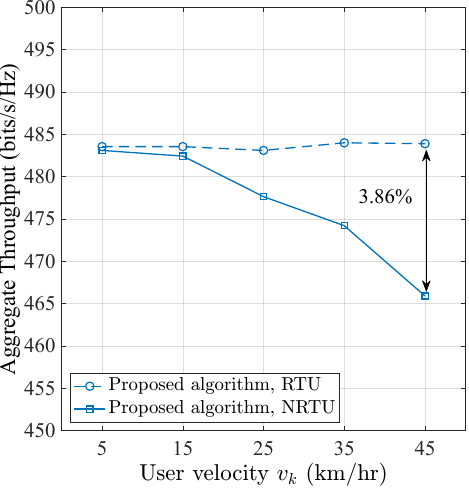}}}
{\subfloat[$N_\mathrm{t}=8$]{\includegraphics[height=45mm]{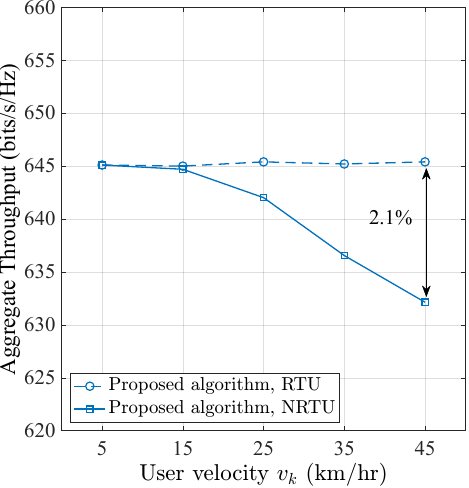}}}
\caption{The aggregate throughput versus user velocity for the proposed algorithm with RT updates (RTU) and near-RT updates (NRTU) of $\mathbf{U}_k$ and $\mathbf{W}_k$.}
\label{fig:speed}
\end{figure}

\begin{rev}
Fig. \ref{fig:speed} shows the aggregate throughput versus user velocity $v_k$ for the proposed framework with RT updates (RTU) and near-RT updates (NRTU) of $\mathbf{U}_k$ and $\mathbf{W}_k$, for the number of transmit antennas $N_\mathrm{t}=4$ and $N_\mathrm{t}=8$. As observed, the performance of the RTU scheme remains nearly constant with increasing user velocity, since the matrices $\mathbf{U}_k$, $\mathbf{W}_k$, and $\mathbf{V}_{k,l}$ are updated in each RT loop. On the other hand, the NRTU scheme exhibits a slight degradation due to less frequent updates of $\mathbf{U}_k$ and $\mathbf{W}_k$. Nevertheless, the performance loss remains limited across all considered velocities. In particular, at $v_k=45$ km/hr, the performance degradation is only $3.86\%$ for $N_\mathrm{t}=4$ and $2.1\%$ for $N_\mathrm{t}=8$. The smaller gap for $N_\mathrm{t}=8$ indicates that increasing the number of antennas per O-RU improves the robustness of the proposed multi-timescale framework. This can be attributed to a stronger channel-hardening effect, which reduces the sensitivity of $\mathbf{U}_k$ and $\mathbf{W}_k$ to channel fluctuations. These results confirm that the proposed framework remains effective under higher user mobility.
\end{rev}

\begin{figure}[t]
\center{\includegraphics[height=50mm]{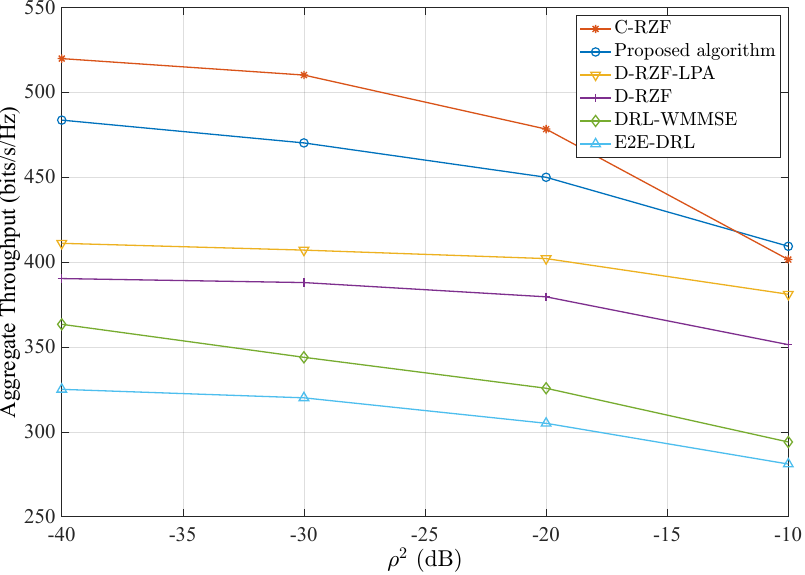}}
\caption{The aggregate throughput versus the channel estimation error.}
\label{fig:CEE}
\end{figure}

Finally, we evaluate the effect of imperfect CSI on the performance of the proposed framework. To do so, we provide the models with a noisy channel estimate given by $\hat{\mathbf{H}}_{k,l}=\mathbf{H}_{k,l} + \widetilde{\mathbf{H}}_{k,l}$. Here, $\widetilde{\mathbf{H}}_{k,l} \in \mathbb{C}^{N_\textrm{r}\times N_\textrm{t}}$ denotes the channel estimation error, where each entry follows a complex Gaussian distribution $\mathcal{CN}(0,\rho^2\beta_{k,l})$. The parameter $\rho^2$ represents the relative CSI error power. Fig. \ref{fig:CEE} shows the aggregate throughput of the proposed framework and baselines for different error levels. As expected, the performance of all algorithms is degraded as $\rho^2$ increases. However, the performance gap between the proposed framework and C-RZF becomes smaller at higher error levels. Notably, at $\rho^2=-10$ dB, the proposed framework outperforms all of the baseline schemes including C-RZF. 
This is because learning-based methods, once trained on noisy data, exhibit robustness to errors in input data when compared with the analytical model-based approaches.

\section{Conclusion} \label{sec:conclusion}
In this paper, we proposed a distributed precoding framework for cell-free massive MIMO within the O-RAN architecture. We formulated a precoding optimization problem to maximize the aggregate throughput while satisfying the minimum rate requirements of users. To solve this nonconvex problem, we reformulated it as an equivalent WMMSE problem and proposed an algorithm to iteratively update the precoding, weight, and receive filter matrices. In order to reduce computational complexity, we used the update equations of the iterative algorithm as expert knowledge to train a multi-agent DRL framework. In each near-RT loop, the DRL agents at the near-RT RIC determine the receive filter and weight matrices for the users. In each RT loop, the O-DUs use the channel matrices along with the latest receive filter and weight matrices received from the near-RT RIC to compute the precoding matrices for their associated O-RUs. Simulation results demonstrated that the proposed framework outperforms distributed baselines by up to $50.2\%$ in terms of the aggregate throughput and performs close to the centralized baseline. The proposed framework also reduces the load on the E2 interface by up to $99.81\%$ compared with the centralized baseline. 
Moreover, it can satisfy the minimum rate requirements of users and dynamically adapt to changes in these requirements. 
For future work, we plan to develop distributed pilot assignment algorithms to further improve the performance of the proposed framework for the case of imperfect CSI.

\appendices
\section{Deriving Subproblem \eqref{eq:WMMSE_V} From Problem \eqref{eq:WMMSE_problem}}
When optimizing for $\mathbf{V}$, the objective function of problem \eqref{eq:WMMSE_problem} can be expressed as
\begin{equation} \label{eq:WMMSE_V_1}
\begin{split}
\sum_{k \in \mathcal{K}} \omega_k \tr \left[ \vphantom{\sum_{i \in \mathcal{K} \setminus \{k\}}} \mathbf{W}_k \left(\mathbf{I}_{N_\textrm{s}} - \mathbf{U}_k^\textrm{H} \mathbf{\Xi}_{k,k}\right) \left(\mathbf{I}_{N_\textrm{s}} - \mathbf{U}_k^\textrm{H} \mathbf{\Xi}_{k,k}\right)^\textrm{H} \right.\\ \left.  + \mathbf{W}_k \sum_{i \in \mathcal{K} \setminus \{k\}}\mathbf{U}_k^\textrm{H} \mathbf{\Xi}_{k,i} \mathbf{\Xi}_{k,i}^\textrm{H} \mathbf{U}_k\right],
\end{split}
\end{equation}
which can further be simplified as
\begin{equation} \label{eq:WMMSE_V_2}
\begin{split}
\minimize_{\mathbf{V} \in \mathcal{D}}\quad \sum_{k \in \mathcal{K}} \omega_k \tr \left[ \sum_{i \in \mathcal{K}} \mathbf{W}_k \mathbf{U}_k^\textrm{H} \mathbf{\Xi}_{k,i} \mathbf{\Xi}_{k,i}^\textrm{H} \mathbf{U}_k \right.\\ \left. -\; \mathbf{W}_k \left(\mathbf{U}_k^\textrm{H} \mathbf{\Xi}_{k,k} + \mathbf{\Xi}_{k,k}^\textrm{H} \mathbf{U}_k\right)\vphantom{\sum_{i \in \mathcal{K}}}\right]. 
\end{split}
\end{equation}
Note that we have $\tr(\mathbf{AB})=\tr(\mathbf{BA})$ \cite{matrix_cookbook}. Thus, problem \eqref{eq:WMMSE_V_2} can be expressed as
\begin{equation} \label{eq:WMMSE_V_3}
\begin{split}
\minimize_{\mathbf{V} \in \mathcal{D}}\quad \sum_{k \in \mathcal{K}} \omega_k \tr \left[ \sum_{i \in \mathcal{K}} \mathbf{\Xi}_{k,i}^\textrm{H} \mathbf{X}_k \mathbf{\Xi}_{k,i} \right.\\ \left. -\; \mathbf{W}_k \mathbf{U}_k^\textrm{H} \mathbf{\Xi}_{k,k} - \mathbf{\Xi}_{k,k}^\textrm{H} \mathbf{U}_k \mathbf{W}_k \vphantom{\sum_{i \in \mathcal{K}}}\right]. 
\end{split}
\end{equation}
The optimal $\mathbf{W}_k$ is a Hermitian matrix according to \eqref{eq:W_WMMSE}. Furthermore, we have $\tr(\mathbf{X} + \mathbf{X}^\textrm{H}) = 2\re(\tr(\mathbf{X}))$. Thus, problem \eqref{eq:WMMSE_V_3} can be expressed as
\begin{equation} \label{eq:WMMSE_V_4}
\begin{split}
\minimize_{\mathbf{V} \in \mathcal{D}}\quad \sum_{k \in \mathcal{K}} \omega_k \tr \left[ \sum_{i \in \mathcal{K}} \mathbf{\Xi}_{k,i}^\textrm{H} \mathbf{X}_k \mathbf{\Xi}_{k,i} - 2\re\left\{\mathbf{Y}_k \mathbf{\Xi}_{k,k}\right\}\right],
\end{split}
\end{equation}
which is equivalent to problem \eqref{eq:WMMSE_V}.

\section{Deriving Subproblem \eqref{eq:WMMSE_ORU} From Problem \eqref{eq:WMMSE_V}}
We first expand the objective function as 
\begin{equation}\label{eq:WMMSE_ORU_1}
    \begin{split}
        \sum_{k \in \mathcal{K}} \omega_k \sum_{i \in \mathcal{K}} \tr \left[ \left( \sum_{l \in \mathcal{L}^\textrm{UE}_i} \mathbf{H}_{k,l} \mathbf{V}_{i,l} \right)^\textrm{H} \mathbf{X}_k \sum_{l \in \mathcal{L}^\textrm{UE}_i} \mathbf{H}_{k,l} \mathbf{V}_{i,l} \right] \\
        - \;2 \sum_{k \in \mathcal{K}} \omega_k \sum_{l \in \mathcal{L}^\textrm{UE}_k} \re \left\{ \tr \left[ \mathbf{Y}_k \mathbf{H}_{k,l} \mathbf{V}_{k,l}\right] \right\},
    \end{split}
\end{equation}
which is equivalent to
\begin{equation}\label{eq:WMMSE_ORU_2}
    \begin{split}
        \sum_{i \in \mathcal{K}} \omega_i \sum_{k \in \mathcal{K}} \tr \left[ \left( \sum_{l \in \mathcal{L}^\textrm{UE}_k} \mathbf{H}_{i,l} \mathbf{V}_{k,l} \right)^\textrm{H} \mathbf{X}_i  \sum_{l \in \mathcal{L}^\textrm{UE}_k} \mathbf{H}_{i,l} \mathbf{V}_{k,l}  \right] \\
        -\; 2 \sum_{l \in \mathcal{L}} \sum_{k \in \mathcal{K}_l} \omega_k  \re \left\{ \tr \left[ \mathbf{Y}_k \mathbf{H}_{k,l} \mathbf{V}_{k,l}\right] \right\}.
    \end{split}
\end{equation}
Note that $\mathbf{X}_k$ is Hermitian according to \eqref{eq:A_definition}. Thus, we can extract the terms that involve the precoding matrices of O-RU $l$ as
\begin{equation}\label{eq:WMMSE_ORU_3}
    \begin{split}
        2 \sum_{i \in \mathcal{K}} \omega_i \sum_{k \in \mathcal{K}_l} \re \left\{ \tr \left[ \mathbf{Z}_{i,k,l}^\textrm{H} \mathbf{X}_i \mathbf{H}_{i,l} \mathbf{V}_{k,l} \right] \right\} \\
        +\; \sum_{i \in \mathcal{K}} \omega_i \sum_{k \in \mathcal{K}_l} \tr \left[ \left( \mathbf{H}_{i,l} \mathbf{V}_{k,l} \right)^\textrm{H} \mathbf{X}_i  \mathbf{H}_{i,l} \mathbf{V}_{k,l}  \right] \\
        -\; 2 \sum_{k \in \mathcal{K}_l} \omega_k \re \left\{ \tr \left[ \mathbf{Y}_k \mathbf{H}_{k,l} \mathbf{V}_{k,l} \right] \right\},
    \end{split}
\end{equation}
which is equivalent to the objective function \eqref{eq:WMMSE_ORU_obj} since $\tr (\mathbf{AB}) = \tr (\mathbf{BA})$ \cite{matrix_cookbook}.
\bibliographystyle{IEEEtran}
\bibliography{Reference}

@article{Bjornson:CF-mMIMO,
  title={Foundations of user-centric cell-free massive {MIMO}},
  author={Demir, {\"O}zlem Tugfe and Bj{\"o}rnson, Emil and Sanguinetti, Luca},
  journal={Found. Trends Signal Process.},
  volume={14},
  number={3-4},
  pages={162--472},
  year={2021}
}

@article{survey:O-RAN1,
  title={Understanding \relax{O-RAN}: Architecture, interfaces, algorithms, security, and research challenges},
  author={Polese, Michele and Bonati, Leonardo and D’Oro, Salvatore and Basagni, Stefano and Melodia, Tommaso},
  journal={IEEE Commun. Surveys \& Tuts.},
  volume={25},
  number={2},
  pages={1376--1411},
  year={2023},
  month={second quarter}
}

@inproceedings{eMBB-URLLC:dist1,
  title={Coexistence of \relax{eMBB} and \relax{URLLC} in open radio access networks: A distributed learning framework},
  author={Alsenwi, Madyan and Lagunas, Eva and Chatzinotas, Symeon},
  booktitle={Proc. IEEE Global Commun. Conf. (GLOBECOM)},
  address={Rio de Janeiro, Brazil},
  year={2022},
  month={Dec.}
}

@article{MIMO1,
  title={An iteratively weighted {MMSE} approach to distributed sum-utility maximization for a {MIMO} interfering broadcast channel},
  author={Shi, Qingjiang and Razaviyayn, Meisam and Luo, Zhi-Quan and He, Chen},
  journal={IEEE Trans. Signal Process.},
  volume={59},
  number={9},
  pages={4331--4340},
  year={2011},
  month={Sep.}
}

@article{MIMO:E2E1,
  title={End-to-End Deep Learning for {TDD MIMO} Systems in the {6G} Upper Midbands},
  author={Park, Juseong and Sohrabi, Foad and Ghosh, Amitava and Andrews, Jeffrey G},
  journal={IEEE Trans. Wireless Commun.},
  volume={24},
  number={3},
  pages={2110--2125},
  year={2025},
  month={Mar.}
}

@article{MIMO:E2E2,
  title={Deep learning for multi-user {MIMO} systems: Joint design of pilot, limited feedback, and precoding},
  author={Jang, Jeonghyeon and Lee, Hoon and Kim, Il-Min and Lee, Inkyu},
  journal={IEEE Trans. Commun.},
  volume={70},
  number={11},
  pages={7279--7293},
  year={2022},
  month={Nov.}
}

@article{MIMO:WMMSE-unfolding,
  title={Iterative algorithm induced deep-unfolding neural networks: Precoding design for multiuser {MIMO} systems},
  author={Hu, Qiyu and Cai, Yunlong and Shi, Qingjiang and Xu, Kaidi and Yu, Guanding and Ding, Zhi},
  journal={IEEE Trans. Wireless Commun.},
  volume={20},
  number={2},
  pages={1394--1410},
  year={2021},
  month={Feb.}
}

@article{MIMO:precoding_DRL2,
  title={Deep Reinforcement Learning for Distributed Dynamic Coordinated Beamforming in Massive {MIMO} Cellular Networks},
  author={Ge, Jungang and Liang, Ying-Chang and Zhang, Liao and Long, Ruizhe and Sun, Sumei},
  journal={IEEE Trans. Wireless Commun.},
  volume={23},
  number={5},
  pages={4155--4169},
  year={2024},
  month={May}
}

@article{MIMO:SAC_QoS,
  title={Soft Actor-Critic-Based Multi-User Multi-{TTI} {MIMO} Precoding in Multi-Modal Real-Time Broadband Communications},
  author={Huang, Yingzhi and Chi, Kaiyi and Yang, Qianqian and Yang, Zhaohui and Zhang, Zhaoyang},
  journal={IEEE Trans. Wireless Commun},
  volume={23},
  number={12},
  pages={18286--18301},
  year={2024},
  month={Dec.}
}

@article{MIMO:pilot1,
  title={A decentralized pilot assignment algorithm for scalable {O-RAN} cell-free massive {MIMO}},
  author={Oh, Myeung Suk and Das, Anindya Bijoy and Hosseinalipour, Seyyedali and Kim, Taejoon and Love, David J and Brinton, Christopher G},
  journal={IEEE J. Sel. Areas Commun.},
  volume={42},
  number={2},
  pages={373--388},
  year={2024},
  month={Feb.}
}

@article{MIMO:clustering2,
  title={Learning-based Precoding-aware Radio Resource Scheduling for Cell-free {mMIMO} Networks},
  author={Girycki, Adam and Rahman, Md Arifur and Vinogradov, Evgenii and Pollin, Sofie},
  journal={IEEE Trans. Wireless Commun.},
  volume={23},
  number={5},
  pages={4876--4888},
  year={2024},
  month={May}
}

@inproceedings{MIMO:precoding_DRL3,
  title={Multi-agent deep reinforcement learning ({MADRL}) meets multi-user {MIMO} systems},
  author={Lee, Heunchul and Jeong, Jaeseong},
  booktitle={Proc. IEEE Global Commun. Conf. (GLOBECOM)},
  address={Madrid, Spain},
  year={2021},
  month={Dec.}
}

@inproceedings{MIMO:MHSH1,
  title={Distributed Precoding for {eMBB} and {URLLC} Traffic in Cell-free {O-RAN}: A Multi-agent Reinforcement Learning Framework},
  author={Shokouhi, Mohammad Hossein and W.S. Wong, Vincent},
  booktitle={Proc. IEEE Int. Conf. Commun. (ICC)},
  address={Montreal, Canada},
  year={2025},
  month={Jun.}
}

@article{MIMO:power1,
  title={Downlink power control for cell-free massive {MIMO} with deep reinforcement learning},
  author={Luo, Lirui and Zhang, Jiayi and Chen, Shuaifei and Zhang, Xiaodan and Ai, Bo and Ng, Derrick Wing Kwan},
  journal={IEEE Trans. Veh. Technol.},
  volume={71},
  number={6},
  pages={6772--6777},
  year={2022},
  month={Jun.}
}

@inproceedings{MIMO:precoding_DRL4,
  title={Deep reinforcement learning for energy-efficient beamforming design in cell-free networks},
  author={Li, Weilai and Ni, Wanli and Tian, Hui and Hua, Meihui},
  booktitle={Proc. IEEE Wireless Commun. Netw. Conf. Workshops (WCNCW)},
  address={Nanjing, China},
  year={2021},
  month={Mar.}
}

@inproceedings{WMMSE:non_coherent,
  title={{WMMSE} beamforming for user-centric cell-free networks with non-coherent joint transmission},
  author={Wang, Xi and Zhao, Xiaotong and Wang, Juncheng and Shi, Qingjiang},
  booktitle={Proc. IEEE Global Commun. Conf. (GLOBECOM)},
  address={Kuala Lumpur, Malaysia},
  year={2023},
  month={Dec.}
}

@article{WMMSE:reduced,
  title={Generalized reduced-{WMMSE} approach for cell-free massive {MIMO} with per-{AP} power constraints},
  author={Yoo, Wonsik and Yu, Daesung and Lee, Hoon and Park, Seok-Hwan},
  journal={IEEE Wireless Commun. Letters},
  volume={13},
  number={10},
  pages={2682--2686},
  year={2024},
  month={Oct.}
}

@article{melodia:dApp_full,
  title={{dApps}: Enabling Real-Time {AI}-Based Open {RAN} Control},
  author={Lacava, Andrea and Bonati, Leonardo and Mohamadi, Niloofar and Gangula, Rajeev and Kaltenberger, Florian and Johari, Pedram and D'Oro, Salvatore and Cuomo, Francesca and Polese, Michele and Melodia, Tommaso},
  journal={Comput. Netw.},
  volume={269},
  pages={1--18},
  year={2025},
  month={Sep.}
}

@inproceedings{melodia:xApp-scheduler,
  title={{TailO-RAN}: {O-RAN} Control on Scheduler Parameters to Tailor {RAN} Performance},
  author={Longhi, Nicolo and D'Oro, Salvatore and Bonati, Leonardo and Polese, Michele and Verdone, Roberto and Melodia, Tommaso},
  booktitle={Proc. IEEE Global Commun. Conf. (GLOBECOM)},
  address={Taipei, Taiwan},
  year={2025},
  month={Dec.}
}

@article{bjornson:O-RAN1,
  title={Cell-free massive \relax{MIMO} in \relax{O-RAN}: Energy-aware joint orchestration of cloud, fronthaul, and radio resources},
  author={Demir, {\"O}zlem Tugfe and Masoudi, Meysam and Bj{\"o}rnson, Emil and Cavdar, Cicek},
  journal={IEEE J. Sel. Areas Commun.},
  volume={42},
  number={2},
  pages={356--372},
  year={2024},
  month={Feb.}
}

@inproceedings{ML:SAC,
  title={Soft actor-critic: Off-policy maximum entropy deep reinforcement learning with a stochastic actor},
  author={Haarnoja, Tuomas and Zhou, Aurick and Abbeel, Pieter and Levine, Sergey},
  booktitle={Proc. Int. Conf. Mach. Learn. (ICML)},
  address={Stockholm, Sweden},
  year={2018},
  month={Jul.},
}

@article{data_rate2:wong-mehrabian,
  title={Joint Spectrum, Precoding, and Phase Shifts Design for {RIS}-Aided Multiuser {MIMO THz} Systems},
  author={Mehrabian, Ali and W.S. Wong, Vincent},
  journal={IEEE Trans. Commun.},
  year={2024},
  month={Aug.},
  volume={72},
  number={8},
  pages={5087--5101}
}

@article{BenchMARL,
  author  = {Matteo Bettini and Amanda Prorok and Vincent Moens},
  title   = {{BenchMARL}: Benchmarking Multi-Agent Reinforcement Learning},
  journal = {J. Mach. Learn. Res.},
  year    = {2024},
  month={Jul.},
  volume  = {25},
  number  = {217},
  pages   = {1--10},
}

@inproceedings{TorchRL,
  title={{TorchRL}: A data-driven decision-making library for {PyTorch}},
  author={Bou, Albert and Bettini, Matteo and Dittert, Sebastian and Kumar, Vikash and Sodhani, Shagun and Yang, Xiaomeng and De Fabritiis, Gianni and Moens, Vincent},
  booktitle={Proc. Int. Conf. Learn. Represent. (ICLR)},
  address={Vienna, Austria},
  year={2024},
  month={May}
}

@manual{matrix_cookbook,
  author       = {Petersen, Kaare Brandt and Pedersen, Michael Syskind},
  title        = {The Matrix Cookbook},
  organization = {Technical University of Denmark},
  year         = {2012},
  url          = {http://www2.compute.dtu.dk/pubdb/pubs/3274-full.html}
}

@misc{O_RAN_description,
  title        = {{O-RAN} Architecture Description},
  author       = {{O-RAN Alliance}},
  month        = {Technical Specification WG1 TS OAD R005 V16.0, Feb.},
  year         = {2026},
}

@misc{O_RAN_split,
  title        = {{O-RAN} Control, User and Synchronization Plane Specification},
  author       = {{O-RAN Alliance}},
  month        = {Technical Specification WG4 TS CUS WG4 R005 V20.0, Feb.},
  year         = {2026},
}

@misc{Near_RT_RIC_Architecture,
  title        = {{O-RAN} Near-{RT} {RIC} Architecture},
  author       = {{O-RAN Alliance}},
  month        = {Technical Specification WG3 TS RICARCH R005 V8.0, Feb.},
  year         = {2026},
}

@misc{3GPP_split_options,
  title        = {Study on New Radio Access Technology: Radio Access Architecture and Interfaces ({Release 14})},
  author       = {{3GPP}},
  month        = {Technical Report TR 38.801 V14.0.0, Mar.},
  year         = {2017},
}

@misc{3GPP_pathloss,
  title        = {Evolved Universal Terrestrial Radio Access ({E-UTRA}); Further Advancements for {E-UTRA} Physical Layer Aspects ({Release 9})},
  author       = {{3GPP}},
  month        = {Technical Report TR 36.814 V9.2.0, Mar.},
  year         = {2017},
}

@book{nonlinear_prog,
  title={Nonlinear Programming},
  author={Bertsekas, Dimitri P},
  year={2016},
  edition={3rd},
  publisher={Athena Scientific}
}

@book{convex,
  title={Convex Optimization},
  author={Boyd, Stephen P and Vandenberghe, Lieven},
  year={2004},
  publisher={Cambridge Univ. Press}
}

@article{CF-MMIMO,
  title={Cell-free massive {MIMO} versus small cells},
  author={Ngo, Hien Quoc and Ashikhmin, Alexei and Yang, Hong and Larsson, Erik G and Marzetta, Thomas L},
  journal={IEEE Trans. Wireless Commun.},
  volume={16},
  number={3},
  pages={1834--1850},
  year={2017},
  month={Mar.}
}

@article{D-RZF-LPA,
  title={Learning-based downlink power allocation in cell-free massive {MIMO} systems},
  author={Zaher, Mahmoud and Demir, {\"O}zlem Tugfe and Bj{\"o}rnson, Emil and Petrova, Marina},
  journal={IEEE Trans. Wireless Commun.},
  volume={22},
  number={1},
  pages={174--188},
  year={2023},
  month={Jan.}
}
\end{document}